\documentclass[useAMS,usenatbib]{mnras}

\usepackage{natbib, aas_macros}

\usepackage{lmodern}
\usepackage{cite}
\usepackage[dvips]{graphicx}
\usepackage{wrapfig}
\graphicspath{{image/}}
\usepackage{color}
\usepackage{amsmath}
\usepackage{amssymb}
\usepackage{rotating}
\usepackage{gensymb}
\usepackage{appendix}
\usepackage{mathrsfs}
\usepackage{multirow}

\bibpunct{(}{)}{;}{a}{}{,} 

\newcommand{\HI}{\rm H~{\sc i }}
\newcommand{\HII}{\rm H~{\sc ii }}
\newcommand{\HeI}{\rm He~{\sc i }}
\newcommand{\HeII}{\rm He~{\sc ii }}

\newcommand{\TB}{\delta T_{\rm b}}
\newcommand{\MSUN}{{\rm M}_{\odot}}
\newcommand{\LSUN}{\rm L_{\odot}}

\newcommand{\XHII}{x_{\rm HII}}

\newcommand{\TS}{T_{\rm S}}
\newcommand{\TK}{T_{\rm K}}
\newcommand{\TCMB}{T_{\gamma}}
\newcommand{\lya}{\rm {Ly{\alpha}}}

\newcommand{\OmegaB}{\Omega_{\rm B}}
\newcommand{\Omegam}{\Omega_{\rm m}}

\def\DEL2{\Delta^2}

\title[Detecting the first sources using SKA]{21-cm signature of the first sources in the Universe: Prospects of detection with SKA}

\author[Ghara, Choudhury \& Datta]{Raghunath Ghara$^1$\thanks{Email: raghunath@ncra.tifr.res.in}, T. Roy Choudhury$^1$\thanks{Email: tirth@ncra.tifr.res.in} and Kanan K. Datta$^{2}$\thanks{Email: kanan.physics@presiuniv.ac.in} \\
$^1$ National Centre for Radio Astrophysics, TIFR, Post Bag 3, Ganeshkhind, Pune 411007, India\\ 
$^2$ Department of Physics, Presidency University, 86/1 College Street, Kolkata - 700073, India } 

\begin{document}

\date{Accepted ?; Received ??; in original form ???}

\pagerange{\pageref{firstpage}--\pageref{lastpage}} \pubyear{?}

\maketitle

\label{firstpage}


\begin{abstract}
Currently several low-frequency experiments are being planned to study the nature of the first stars using the redshifted 21-cm signal from the cosmic dawn and epoch of reionization. Using a one-dimensional radiative transfer code, we model the 21-cm signal pattern around the early sources for different source models, i.e., the metal-free Population III (PopIII) stars, primordial galaxies consisting of Population II (PopII) stars, mini-QSOs and high-mass X-ray binaries (HMXBs). We investigate the detectability of these sources by comparing the 21-cm visibility signal with the system noise appropriate for a telescope like the SKA1-low. Upon integrating the visibility around a typical source over all baselines and over a frequency interval of 16 MHz, we find that it will be possible to make a $\sim 9-\sigma$ detection of the isolated sources like PopII galaxies, mini-QSOs and HMXBs at $z \sim 15$ with the SKA1-low in 1000 hours. The exact value of the signal to noise ratio (SNR) will depend on the source properties, in particular on the mass and age of the source and the escape fraction of ionizing photons. The predicted SNR decreases with increasing redshift. We provide simple scaling laws to estimate the SNR for different values of the parameters which characterize the source and the surrounding medium. We also argue that it will be possible to achieve a SNR $\sim 9$ even in the presence of the astrophysical foregrounds by subtracting out the frequency-independent component of the observed signal. These calculations will be useful in planning 21-cm observations to detect the first sources.

\end{abstract}


\section{Introduction}
\label{intro}

In recent times, a large number of galaxies at  $z > 6$ have been detected using narrow-band $\lya$ emission \citep[e.g.,][]{Ouchi10, Hu10, Kashikawa11} and broad-band colour \citep{Ellis13, Bouwens15}. In addition, various surveys have been used to detect tens of quasars at $6 \lesssim z \lesssim 7$ \citep{Fan06b, Mortlock11, Venemans15}. These sources are believed to play an important role during the last phase of the epoch of reionization (EoR). As a next step, it is natural to ask the question whether one can detect the very first galaxies in the Universe using similar techniques.  In spite of significant progress in theoretical modelling of the first sources \citep{thomas08, 2014MNRAS.445.3674Y}, there are still no observational signatures. It is expected that future space missions like the James Webb Space Telescope (JWST)\footnote{http://jwst.nasa.gov} will be able to detect at least the brightest of these sources \citep{2011ApJ...740...13Z, 2013MNRAS.436.1555D, 2014MNRAS.442.1640D}. It is believed  that the UV photons from these sources created ionized bubbles around them, which eventually overlapped and completed the reionization process by redshift $\sim 6$ \citep{fan2006, Choudhury06a,  Goto11, mitra2011, mitra2012, Mitra15}. The presence of these bubbles motivated various groups to explore an alternate, perhaps somewhat indirect, method of detecting the high-redshift sources. It has been suggested that one can detect the 21-cm signatures around these sources using low-frequency telescopes \citep{kanan2007MNRAS.382..809D, 2008MNRAS.386.1683G, Datta2012b} which can be helpful in constraining the source properties \citep{2012MNRAS.426.3178M}.

According to our current understanding, the first stars, usually known as the Population III (PopIII) stars, formed in a metal free (or low metallicity) environment and thus are expected to be much more massive than the present day stars \citep{2002Sci...295...93A, 2007ApJ...654...66O, 2009Natur.459...49B, wise2012}. A population of these stars would have caused supernova explosions, which would contaminate the surrounding medium with metals. The resulting chemical feedback would result in formation of the Population II (PopII) stars. In addition to the stellar population, one expects the galaxies to host X-ray sources like the mini-QSOs, the hot interstellar gas or the high-mass X-ray binaries (HMXBs). These X-ray sources are important in increasing the kinetic temperature of the gas in the intergalactic medium (IGM) above the brightness temperature of the cosmic microwave background radiation (CMBR). All such source properties are imprinted in the redshifted 21-cm signal from the neutral hydrogen from these epochs. Analytical calculations \citep[e.g.,][]{furlanetto04, 2014MNRAS.442.1470P}, semi-numerical simulations \citep{zahn2007, mesinger07, santos08, Thom09, choudhury09, ghara15a, ghara15b}, and full numerical simulations involving radiative transfer \citep{Iliev2006,  mellema06, McQuinn2007, shin2008, baek09}  have been carried out to understand the effect of different sources on the global and statistical quantities (e.g., the power spectrum) of the signal.  Relatively less attention has been paid to understand the nature of the 21-cm signal around individual sources, though there have been studies which show that the 21-cm structure around the different sources depend on the nature of spectral energy distribution (SED) of the sources and IGM properties \citep{thomas08, Alvarez10, 2012MNRAS.426.3178M, 2012MNRAS.425.2964Z, 2014MNRAS.445.3674Y}. 

Though the high redshift 21-cm signal is expected to carry information about the first sources, it is quite challenging to detect the signal and extract this information. The main difficulty is that the strength of the cosmological signal is very weak compared to the typical system noise and the foregrounds. The system noise increases at low frequencies and hence one requires optimal baseline design and large observation times to keep the noise below the expected signal. While the first generation of low-frequency radio telescopes like the Low Frequency Array (LOFAR)\footnote{http://www.lofar.org/} \citep{van13}, the Precision Array for Probing the Epoch of Reionization (PAPER)\footnote{http://eor.berkeley.edu/} \citep{parsons13}, the Murchison Widefield Array (MWA)\footnote{http://www.mwatelescope.org/} \citep{bowman13, tingay13}, and the Giant Metrewave Radio Telescope (GMRT)\footnote{http://www.gmrt.tifr.res.in}\citep{ghosh12, paciga13} are engaged in detecting the signal statistically at $z < 12$, a highly sensitive telescope like the Square Kilometre Array (SKA)\footnote{http://www.skatelescope.org/} will have much better sensitivity at lower frequencies allowing us to probe even higher redshifts. 

Given that the signal from the first sources is very weak, the first task would be to simply make a detection of the signatures of these early sources in low-frequency observations. Once the detection is confirmed, one can follow it up and make further progress by constraining various properties of these sources. It would thus be interesting to explore the detectability of these first sources with telescopes like the SKA1-low within reasonable observation time. Keeping this in mind, we have modelled the 21-cm signal around the first sources with the source properties and the properties of the surrounding medium being characterized by a number of parameters. The main goal of this study is to characterize the detectability of the 21-cm signal as a function of these parameters. This would help in planning observations using the SKA1-low so as to make the detection of the signal around the first sources. Similar studies have been done, e.g., by \citet{kanan2007MNRAS.382..809D, 2008MNRAS.386.1683G, Datta2012b}, who showed that the large ionized bubbles around individual sources can be detected using telescopes like the GMRT, LOFAR and the MWA within reasonable integration time around redshift $\sim 8$. In our case, however, the situation is much more complex as we are interested in the very early stages of reionization (i.e., the cosmic dawn) where the IGM contains both emission and absorption regions.

We have organized the paper in the following way. In section \ref{simu}, we calculate the expected 21-cm pattern for different source models and also obtain the expected visibilities and noise for typical observations. The main results of our work are presented in section \ref{res}, before we summarize our findings in section \ref{conc}. Throughout the paper, we have used the cosmological parameters  $\Omegam=0.32$, $\Omega_\Lambda=0.68$, $\OmegaB=0.049$, $h=0.67$, $n_{\rm s}=0.96$, and $\sigma_8=0.83$, which are consistent with the recent results of $Planck$ mission \citep{Planck2013}.

\section{Calculation of the signal}
\label{simu}

\subsection{Isolated radiation source}
\label{source_pro}

In general, we expect the heating and ionization signatures around an individual source to be complicated by the presence of other sources nearby. The signal pattern of other sources would tend to overlap with the target source under consideration, thus making the analysis complicated. However, a large amount of information regarding the detectability of the signal can be extracted by studying the pattern around an isolated source, i.e., ignoring the effect of overlap from other sources. A large fraction of this paper is devoted to understanding the signal properties around an individual isolated source. We will study a more realistic scenario later in Section \ref{over_sou}.

\begin{figure}
\begin{center}
\includegraphics[scale=0.7]{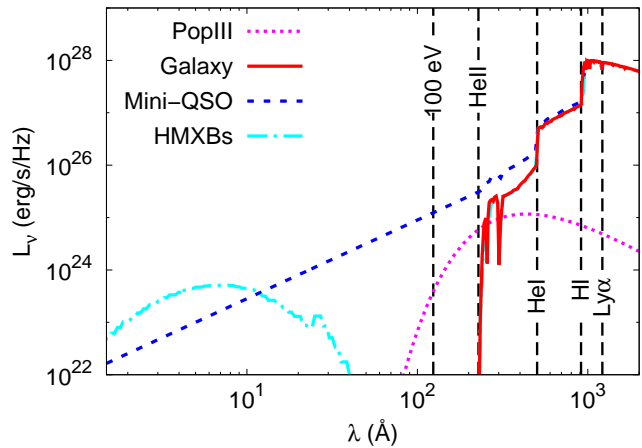}
    \caption{The SED of the source models considered in this work, i.e.,  PopIII, Galaxy, Mini-QSO and HMXBs. The mass of the PopIII star is taken to be $ 10^3\, \MSUN$. The stellar mass in the three models Galaxy, Mini-QSO and HMXBs is taken as $M_{\star} =  10^7\, \MSUN$. The sources are assumed to form in a metal poor environment in the IGM with metallicity $10^{-3}\, Z_\odot$. The ratio of X-ray to UV luminosity ($f_X$) for the Mini-QSO and HMXBs models are fixed to 0.05. We have fixed the power-law index $\alpha = 1.5$ for the Mini-QSO model. The vertical lines from right to left represent the wavelengths corresponding to $\lya$ (10.2 eV), hydrogen ionization energy (13.6 eV), ionization energy of \HeI  (24.6 eV),  ionization energy of \HeII (54.4 eV) and 100 eV respectively. }
   \label{spec}
\end{center}
\end{figure}

Since the exact nature of the first sources of radiation is still unknown, we consider a variety of source models as listed below. The spectral energy distribution (SED) of these sources are shown in Fig \ref{spec}.

\begin{itemize}

\item {\bf PopIII:} In many models of reionization \citep{wl03, Furlanetto06a, Choudhury07, chen2008}, the first stars are expected to be massive since they form in a metal free environment. These are commonly known as the PopIII stars.  In this work, we have modelled radiation from these sources as blackbody spectrum with an effective temperature given by \citep{Bromm01},
\begin{equation}
T_{\rm eff}(M_{\star}) = 1.1 \times 10^{5} \left(\frac{M_{\star}}{100~ \MSUN} \right)^{0.025} \rm K, 
\label{equ_teff}
\end{equation}
where $M_{\star}$ is the mass of the star. The spectral energy distribution (SED) of a PopIII star is normalized by the bolometric luminosity corresponding to the star, which can be written as,
\begin{equation}
L_{\rm bol}(M_{\star}) =  10^{4.5} \frac{M_{\star}}{\MSUN} \LSUN, 
\label{equ_bolo}
\end{equation}
where $\LSUN=3.846\times 10^{33} \rm erg ~s^{-1}$ is the solar luminosity. We assume that a galaxy hosts only one PopIII star, however, our analysis would be valid also for a cluster of several PopIII stars which have a blackbody like SED. We also assume that all photons produced within a PopIII hosting galaxy escape into the IGM.

\item {\bf Galaxy:} The effectiveness of PopIII stars in ionizing and heating the IGM is somewhat uncertain as the gas within galaxies can get polluted by metals after the first burst of star formation. Hence, we have taken an alternate source model which consists of normal (PopII) stars. We have generated the SED of these model galaxies using the stellar population synthesis code {\sc pegase2}\footnote{\tt http://www2.iap.fr/pegase/} \citep{Fioc97} for standard star formation scenarios. We assume that the galaxies form in a metal poor environment with metallicity $10^{-3}\, Z_\odot$, where $Z_\odot$ is solar metallicity and  follow a Salpeter IMF for stars with mass between 1 to 100 $\MSUN$.\footnote{The stars in the primordial galaxies are expected to form in metal poor environments \citep{Lai07,Finkelstein09b}. The transition from PopIII to PopII stars occurs when the metallicity approaches some critical value $Z_{\rm crit}$, which is typically in the range $10^{-6} - 10^{-3}\, Z_\odot$ \citep{2001MNRAS.328..969B, 2010MNRAS.407.1003M}. Hence we make a conservative assumption that the PopII stars form in an environment with metallicity  $10^{-3}\, Z_\odot$.} The main difference of this model compared to PopIII stars is that these galaxies do not produce very high energy photons (i.e., larger than 50 eV, see Fig. \ref{spec}). Also, one should keep in mind that a substantial fraction of ionizing photons produced is absorbed within the host galaxy, only a fraction $f_{\rm esc}$ is assumed to escape into the IGM and contribute to reionization.

\item {\bf Mini-QSO:}  An important component of the SED at high redshifts is the X-ray radiation which plays an important role in heating the IGM. It is possible that the galaxies, in addition to the stellar component, host mini-QSOs that are powered by intermediate mass accreting black holes of mass $10^3-10^6\, \MSUN$. For such sources, we assume that the SED has a stellar component as discussed in the previous paragraph and also a high energy component that follows a power-law with a spectral index $\alpha$  \citep{vanden01, vignali03},

\begin{equation}
I_q(E) = A ~ E^{- \alpha}. 
\label{equ_powerlaw}
\end{equation}

The normalization factor $A$ can be calculated in terms of the UV luminosity\footnote{We have assumed that the UV band span from 10.2 to 100 eV and the X-ray band span from 100 eV to 10 keV.}, the X-ray to UV luminosity fraction $f_X$ and the spectral index $\alpha$. Unlike the previous two source models,  these mini-QSOs emit large number of soft X-ray photons, which are very effective in partially ionizing and heating the surrounding neutral medium \citep{shull1985}. We should mention that the power-law SED represents not only the mini-QSO type sources, but also other sources of X-rays like supernovae,  hot inter-stellar medium etc \citep{Pacucci2014}.

\item {\bf HMXBs:} A different source of X-ray photons could be the HMXBs within galaxies. These sources are different from the mini-QSOs because the soft X-rays in this case will be substantially absorbed due to inter-stellar absorption (as shown in Figure \ref{spec}) and thus the spectrum can not be fitted with any power-law in X-ray band. The level of interstellar absorption of the soft X-ray photons is quite uncertain for the high redshift HMXBs. We have used the SED given by \citet{Frag1, Frag2} and we have assumed that the SED shape is independent of other galaxy properties, i.e., the stellar mass, age of the galaxies, etc. It is possible that the SED of the HMXBs may change with the mean stellar population age because of the evolution in the metallicity \citep{Frag2, 2014MNRAS.440L..26K}. In order to keep our model simple, we assume that the metallicity does not evolve significantly over the time-scale which correspond to the age of the source.

\end{itemize}

The fiducial values of the parameters used in this study are chosen as follows: The fiducial stellar mass is taken to be $M_\star = 10^3\, \MSUN$ for the PopIII model, which can either be a single star or a cluster of several PopIII stars with mass $M_\star \sim 10^2\, \MSUN$. For the other three source models the fiducial stellar mass is chosen to be $M_\star = 10^7\, \MSUN$ which corresponds to stars in dark matter haloes of mass $M_{\rm halo} \sim 6 \times 10^8\, \MSUN$ (assuming a fraction $f_{\star} = 0.1$ of the baryons to convert into stars). The value of the escape fraction is taken to be $f_{\rm esc}=0.1$ for all these sources. We have assumed a power-law spectrum for the Mini-QSO source model with a fiducial spectral index $\alpha = 1.5$, consistent with the observational constraints from \citet{laor97, vanden01, vignali03}. The X-ray to UV luminosity $f_X$ of the Mini-QSO and HMXBs models can be related to the ratio of the mass of the intermediate mass black hole to the galaxy mass. Recent observations of high-$z$ quasars predict  the accretion rate of the black holes (BH) to be similar to the Eddington limit (e.g., \citealt{Willott10b}). We choose the fiducial value of the parameter $f_X$ to be 0.05, which corresponds to a BH to galaxy mass ratio $\sim 10^{-3}$, consistent with observations (e.g., \citealt{Rix04}). Very little is known about the lifetime of the first sources. It is believed that they are short-lived \citep{Meyn05}, hence we choose the fiducial age $t_{\rm age}$ of the sources as 20 Myr. The densities of hydrogen and helium in the IGM are assumed to be uniform and the density contrast $\delta$ is set to 0.


\subsection{Heating and ionization maps around the source}
\label{1drt}

For each source model, we generate the ionization and heating maps around an isolated source using an one dimensional radiative transfer code. We use the method developed in our earlier works \citep{ghara15a, ghara15b} which is based on \citet{thomas08}. The main features of the method used in this work are as follows.

\begin{itemize}
\item The procedure is simplified by assuming a constant density of hydrogen and helium in the IGM surrounding the source. The initial state (i.e., at the instant when the source produces the first radiation) of the IGM is taken to be completely neutral and the initial temperature is computed assuming that the gas temperature evolves as $(1+z)^{2}$ after decoupling from the radiation temperature around redshifts $\sim$ 150.

\item We divide the IGM into several spherical shells along the radial direction from the centre of the source. We assume that the radiation from the source can influence the IGM up to a distance `$c\times t$' from the centre of the source when the age of the source is `$t$', with `$c$' being the speed of light.

\item The intensities at UV and X-ray bands will decrease due to absorption in the medium around the source. We pre-calculate the ionization and heating rates for different values of optical depths. These are later used for calculating the time evolution of ionized species of hydrogen and helium along with the kinetic temperature ($T_{\rm K}$) of the medium.

\item The physical processes that affect the population of the ionization states of hydrogen and helium are photo-ionization by the continuum photons, secondary ionization by high-energy primary electrons emitted due to photo ionization, collisional ionization and recombination. The temperature of the IGM is set by different heating and cooling mechanisms like photo-heating, Compton heating, energy loss due to recombination, collisional excitation etc.

\item The $\lya$ photons are assumed to be contributed by the continuum spectrum of the source, recombination of hydrogen in the ionized ISM and collisional excitation due to secondary electrons. We do not solve a detailed radiative transfer for the $\lya$ photons in this work. We have instead assumed that the $\lya$ flux decreases  as $1/R^2$, where $R$ is the radial distance from the centre of the source.

\item In the initial parts of this study, we have not incorporated the effect of gas velocities on the brightness temperature calculation. In reality, the situation may be complex and one needs to incorporate the density fluctuations and velocity effects while generating the 21-cm signal. We have addressed such complex scenario in Section \ref{over_sou}.

\item In order to understand the characteristics of the signal at various epochs, we consider three different models of heating and $\lya$ coupling, namely:

(i) {\bf Model $A$:} We assume that the IGM is highly heated and $\lya$ coupled, i.e., the spin temperature $\TS >> \TCMB$, where $\TCMB$ is the CMBR brightness temperature at redshift $z$. 

(ii) {\bf Model $B$:} We assume high $\lya$ coupling in the IGM ($\TS = \TK$), however we calculate the $\TK$ distribution around the sources self-consistently.

(iii) {\bf Model $C$:} The $\TS$ profile around the source depends on the source properties and is generated self-consistently.

Model $C$ is probably the most realistic model for the very first sources in the universe as the number density of these sources is small and thus no significant overlap between the signal pattern from different sources is expected. Once the $\lya$ coupling becomes efficient during the very early stages of reionization, one expects model $B$ to represent the IGM conditions. In presence of X-ray sources, e.g., mini-QSOs, a major fraction of the universe is heated above $\TCMB$ once the universe becomes $\gtrsim 10 \%$ ionized and thus model $A$ is expected to hold thereafter \citep{ghara15a}.

\end{itemize}


\subsection{The 21-cm signal}
\label{21_cm}

Once the ionization and heating maps are generated around the source, it is straightforward to calculate the collisional, radiative and $\lya$ coupling coefficients which can then be used for calculating the spin temperature $\TS$. The quantity which is most relevant for observations is the differential brightness temperature $\delta T_b (\vec{\theta}, \nu)$, where $\vec{\theta}$ is a two-dimensional vector on the sky plane characterizing the sky position and $\nu$ is the frequency of observation. It is simply the difference between the 21-cm and the CMBR brightness temperatures, and can be expressed as
\begin{eqnarray}
 \delta T_b (\vec{\theta}, \nu)  \!\!\!\! & = & \!\!\!\! 27 ~ x_{\rm HI} (\mathbf{x}, z) [1+\delta_{\rm B}(\mathbf{x}, z)] \left(\frac{\OmegaB h^2}{0.023}\right) \nonumber\\
&\times& \!\!\!\!\left(\frac{0.15}{\Omegam h^2}\frac{1+z}{10}\right)^{1/2}\left[1-\frac{\TCMB(z)}{T_{\rm S}(\mathbf{x}, z)}\right]\,\rm{mK},
\nonumber \\
\label{brightnessT}
\end{eqnarray}
where the quantities $x_{\rm HI}(\mathbf{x}, z)$ and $\delta_{\rm B}(\mathbf{x}, z)$ denote the neutral hydrogen fraction and the baryonic density contrast respectively at the comoving coordinate $\mathbf{x}$ at a redshift $z=1420~{\rm MHz}/\nu -1$. The three-dimensional position $\mathbf{x}$, if measured from the location of the observer at $z=0$, is related to the sky position $\vec{\theta}$ by the relation $\mathbf{x} = \left\{r(z) \vec{\theta},~ r(z)\right\}$, where $r(z)$ is the comoving distance to $z$, respectively. The quantity $\TCMB(z)$ = 2.73 $\times (1+z)$ K is the CMBR brightness temperature at redshift $z$. We have not considered the effect of the peculiar velocities of the gas in the IGM in the above equation.

We have generated the expected signal in a rectangular box with length along the frequency axis corresponding to the observational bandwidth considered in this work. For example,  a bandwidth of 16 MHz at redshift 15 corresponds to a  simulation box of length 357 comoving Mpc (cMpc) along the corresponding axis.


\subsection{Signal visibilities}
\label{ska_sen}

The measurable quantity in a radio interferometric observation is the visibility. In general, observations are carried out over many frequency channels with frequency resolution $\Delta \nu_c$ over a bandwidth $B_\nu$. The visibility is measured for each frequency channel and for each antenna pair. Thus the visibility $V(\vec{U},\nu)$ is a function of the channel frequency $\nu$ and baseline $\vec{U}=\vec{d}_{\rm ant}/\lambda$, where $\vec{d}_{\rm ant}$ is the two-dimensional separation vector between the antenna pair and $\lambda$ is the wavelength of the observation. Under the approximation that either the antennas are confined to a two-dimensional plane or the field of view (FOV) is small, the visibility can be related to the two-dimensional Fourier transform of the sky intensity distribution, i.e.,
\begin{equation}
V(\vec{U}, \nu) = \int d^2\theta ~I_{\nu}(\vec{\theta})~A(\vec{\theta})~e^{i2\pi \vec{\theta} \cdot \vec{U}},
\label{visibi}
\end{equation} 
where $I_{\nu}(\vec{\theta})$ is the sky specific intensity at frequency $\nu$ and $A(\vec{\theta})$ is the primary beam pattern for individual antenna, which can be assumed to be Gaussian $A(\vec{\theta})= e^{-\theta^2/\theta_{0}^{2}}$ with $\theta_{0} \approx 0.6~ \theta_{\rm FWHM}$. For relatively wide field beam pattern of the antenna, the beam pattern can be taken out from the integral as $A(\vec{\theta_{c}})$, where $\vec{\theta_{c}}$ is the position of the centre of the source. For the rest of the paper, we shall assume that $A(\vec{\theta_{c}}) \approx 1$. This assumption works quite well when the 21-cm emitting region is confined to a small angular extent around the centre of the FOV, e.g., for an isolated source with a short lifetime. The sky intensity can be related to the differential brightness temperature as
\begin{equation}
I_{\nu}(\vec{\theta}) = \frac{2 k_B \nu^2  }{c^2} \delta T_b (\vec{\theta}, \nu),
\label{inten}
\end{equation}
where $k_B$ is Boltzmann constant.

In any observation, the total signal measured by the telescope will include, in addition to the cosmological signal of interest, the inherent system noise and the foreground from the astrophysical sources. In the case of 21-cm observations, these can create enormous difficulties as the cosmological signal is very weak compared to all other sources that contribute to the visibility. In addition there could be man made radio frequency interference (RFI). For simplicity, we assume that the RFI can be identified and the corrupted data can be discarded. We will also assume that the astrophysical foregrounds can be modelled accurately and subtracted from the data\footnote{Complete subtraction of the foregrounds without affecting the cosmic signal and system noise is an overly idealistic assumption. In principle, some amount of signal and system noise will be lost while subtracting the foregrounds. In addition some residual foregrounds would remain as contaminant within the signal, which in turn may affect to the detectability.  We have discussed the detectability of the signal in presence of foregrounds in Section \ref{fore_deal}}. Under such assumptions, we need to account for only the system noise, which too can be quite large at low frequencies. We then write the measured visibility as
\begin{equation}
V(\vec{U}, \nu) = S(\vec{U}, \nu) + N(\vec{U}, \nu),
\label{visi_con}
\end{equation}
where $S(\vec{U}, \nu)$ is the 21-cm signal visibility given by equations (\ref{visibi}) and (\ref{inten}) and $N(\vec{U}, \nu)$ is the system noise contribution to the measured visibility.

The system noise $N(\vec{U}, \nu)$ in different baselines and  frequency channels are uncorrelated and are expected to be Gaussian random variables with zero mean. The rms noise $\sigma_N$ for each  baseline, polarization and frequency channel of width $\Delta \nu_c$ and correlator integration time $\Delta t_{c}$ is given by \citep{thompson},
\begin{equation}
\sqrt{\left< N^2 \right>} = \frac{\sqrt{2} k_B T_{\rm sys}}{A_{\rm eff} \sqrt{\Delta \nu_{c} ~ \Delta t_{c}}},
\label{rms_noise}
\end{equation}
where $A_{\rm eff}$ is the effective collecting area of each antenna and $T_{\rm sys}$ is the system temperature. Note that $\left< N^2 \right>$ is independent of $U$, however it depends on the frequency implicitly through the quantity $T_{\rm sys}$.

\begin{table}
\centering
\small
\tabcolsep 3pt
\renewcommand\arraystretch{1.2}
   \begin{tabular}{ll}
\hline
\hline
    Parameters   & Values \\
\hline
\hline
                Redshift ($z$)                                               &          15                  \\
		Central frequency ($\nu_c$) 				&		$88.75$ MHz	\\
		Band width ($B_{\nu}$)				&		16 MHz							\\
		Frequency resolution ($\Delta \nu_c$)		&		50 kHz							\\
    Observational time	($t_{\rm obs}$)	& 	1000 h					    \\
	System temperature	($T_{\rm sys}$)				&		$60 \times (300 ~\rm MHz / \nu_c)^{2.55} ~\rm K$ \\
		Number of antennas ($N_{\rm ant}$) 	& 	512 (SKA), 30 (GMRT), \\ & 128 (MWA), 48 (LOFAR)		\\					\\
	Effective collecting area ($A_{\rm eff}$)				&		962 $\rm m^2$ (SKA), 1590 $\rm m^2$ (GMRT), \\ & 16 $\rm m^2$ (MWA),  526 	$\rm m^2$ (LOFAR)\\

\hline
\end{tabular}
\caption[]{The SKA1-low, GMRT, MWA and LOFAR parameters used in this study for observations at redshift $z$.}
\label{tab1}
\end{table} 

\begin{figure*}
\begin{center}
\includegraphics[scale=0.7]{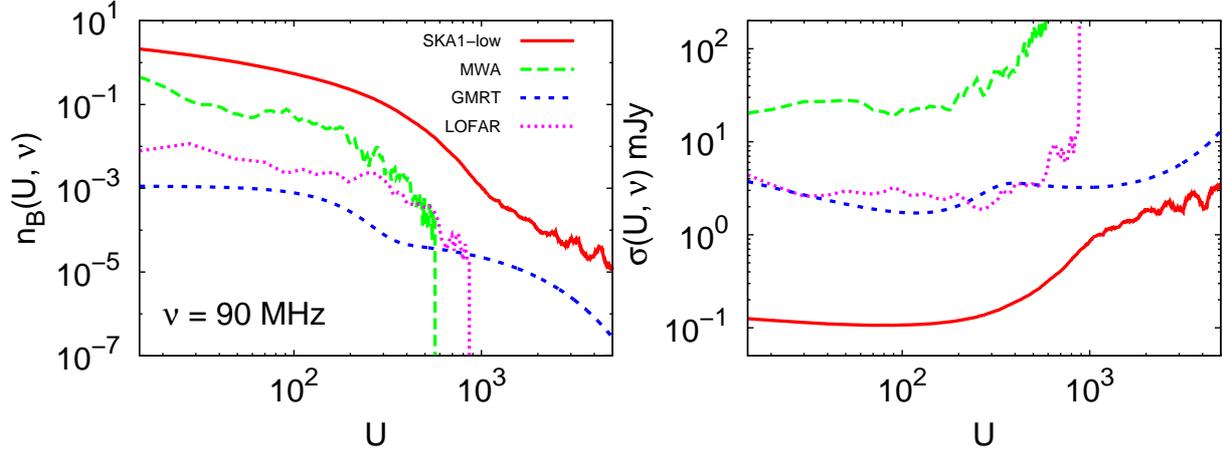}
    \caption{The baseline distributions (left panel) and the corresponding rms noise in visibilities (right panel) for the SKA1-low, MWA, GMRT and LOFAR at frequency 90 MHz. $n_{\rm B}(U,\nu)$ denote the number of antenna pairs having same baseline $U$ at frequency $\nu$. The rms is computed for an observation time of  1000 hours and frequency resolution of 50 kHz. We have fixed $\Delta U$=10 while calculating the rms noise using equation (\ref{reduce_rms}).}
   \label{ska_base}
\end{center}
\end{figure*}

Table \ref{tab1} shows the antenna parameters for various telescopes of interest, i.e., SKA1-low, GMRT, MWA and LOFAR, along with the default observational criteria used in this study. At low frequencies, the system temperature is dominated by the sky temperature which is taken to be $T_{\rm sys} = 60 \times (300 ~\rm MHz / \nu_c)^{2.55} ~\rm K$ \citep{Jensen13, Dewdney2013}. The frequency channel width is taken to be 50 kHz, which corresponds to a spatial resolution of $\sim$ 1.1 cMpc at redshift 15. 

There are various ways of reducing the rms noise in observations. An obvious option is to average over long observation time $t_{\rm obs}$ which reduces the noise by a factor of $\sqrt{\Delta t_{c}/t_{\rm obs}}$. Also it is possible that different combinations of antenna pairs can give same baseline $\vec{U}$ for the same frequency channel. Assuming the antenna distribution to have circular symmetry, the rms noise can be further reduced by a factor of 1/$\sqrt{2\pi n_{\rm B}(U, \nu) U \Delta U}$, where  $2 \pi n_{\rm B}(U, \nu) U \Delta U$ is the number of pair of antennas that have baseline lengths between $U$ and $U + \Delta U$ at a frequency $\nu$.  We denote this reduced rms as $\sigma(U, \nu)$ which can be written as
\begin{equation}
\sigma(U, \nu) =  \frac{\sqrt{2}~ k_B T_{\rm sys}}{A_{\rm eff} \sqrt{\Delta \nu_{c} ~  t_{\rm obs}}}\times \frac{1}{\sqrt{2\pi n_{\rm B}(U, \nu)U \Delta U}}.
\label{reduce_rms}
\end{equation}
The quantity $n_{\rm B}(U, \nu)$ is normalized such that 
\begin{equation}
\int n_{\rm B}(U,\nu) ~d^2U = \frac{N_{\rm ant}(N_{\rm ant}-1)}{2},
\label{norm_base}
\end{equation}
where $N_{\rm ant}$ denotes the number of antennas in the system such that the total number of baselines is given by the right hand side of the above equation. The baseline distributions $n_{\rm B}(U,\nu)$ for various telescopes (GMRT, LOFAR, MWA and SKA1-low) are plotted in the left panel of Figure \ref{ska_base}.\footnote{ We have used the previously proposed form of the SKA baseline distribution for 1024 antennas, but normalized the distribution for 512 antenna using equation (\ref{norm_base}). This baseline distribution is not significantly different from the recently finalized baseline distribution of the SKA1-low as given in http://astronomers.skatelescope.org/wp-content/uploads/2015/11/SKA1-Low-Configuration$\_$V4a.pdf. For GMRT baseline distribution, we have used the fitting function of \citet{kanan2007MNRAS.382..809D}.} The right hand panel shows the corresponding rms defined in equation (\ref{reduce_rms}) as a function of baseline $U$. We can see that, for same amount of observing time, the rms for SKA1-low at smaller baselines $U \lesssim 100$ is at least 10 times better than any other existing facilities. This is because of the significantly greater number of antennas in the core of SKA1-low. As we will see later, the signal we are studying in this paper is mostly concentrated at $U \lesssim 100$, and its amplitude is such that only SKA1-low will be sensitive enough to make a detection. Hence, we will concentrate our discussions mainly on SKA1-low in the rest of the paper.

\begin{figure*}
\begin{center}
\includegraphics[scale=0.6]{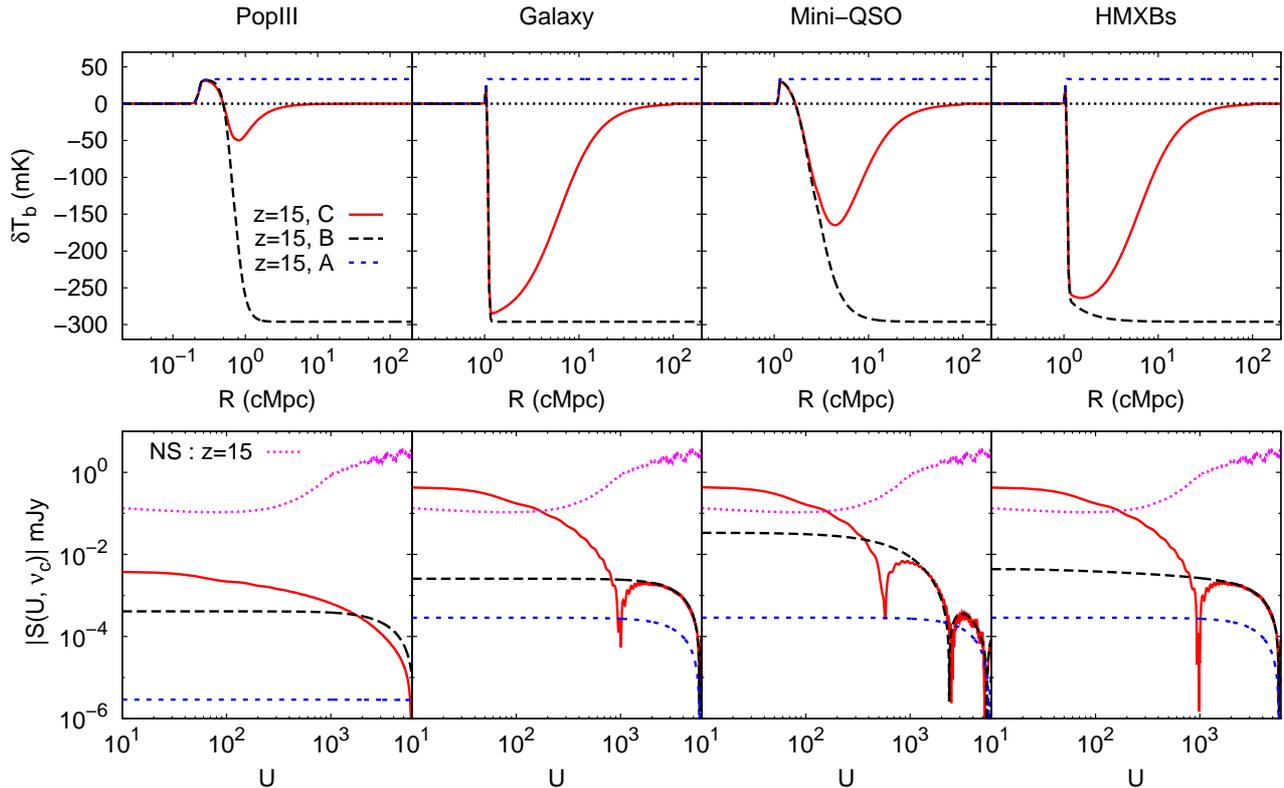}
    \caption{Upper panel: The brightness temperature distribution around an isolated source for different source models, i.e., PopIII, Galaxy, Mini-QSO and HMXBs. The source properties are taken to be those corresponding to the fiducial values. The results are shown for all three coupling models.  Bottom panel: The absolute value of the corresponding visibility amplitude as a function of baseline $U$. Also shown are rms noise in the visibilities calculated for 1000 h of observation with the SKA1-low with a frequency resolution of 50 kHz.  We have fixed $\Delta U$=10 while calculating the rms noise using equation (\ref{reduce_rms}).}
   \label{visi_red}
\end{center}
\end{figure*}

One way of increasing the possibility of detecting the faint 21-cm signal is to integrate it over a wide range of baselines and frequency channels. We define an estimator $\hat{E}$ as
\begin{equation}
\hat{E} = A_{\rm NS} \int {\rm d}^2 U \int {\rm d} \nu~n_B(U,\nu)~V(\vec{U}, \nu),
\label{estimator-1}
\end{equation}
where $A_{\rm NS}$ is a normalization constant given by
\begin{equation}
A_{\rm NS}^{-1} = \int {\rm d}^2 U \int {\rm d} \nu~n_B(U,\nu) = \frac{N_{\rm ant}(N_{\rm ant} - 1)}{2} B_{\nu}.
\end{equation}
Since the noise can be treated as random numbers with zero mean, the expectation value of the estimator is given by
\begin{equation}
\langle \hat{E} \rangle = A_{\rm NS} \int {\rm d}^2 U \int {\rm d} \nu~n_B(U,\nu)~S(\vec{U}, \nu).
\end{equation}
The contribution of the noise term will be included in the standard deviation
\begin{equation}
\sqrt{\left\langle \left(\Delta \hat{E}\right)^2 \right\rangle} \equiv \sigma_N = \frac{\sqrt{2}~ k_B T_{\rm sys}} {A_{\rm eff}~\sqrt{t_{\rm obs}~B_{\nu}~N_{\rm ant} (N_{\rm ant} - 1) / 2}},
\label{snr_n}
\end{equation}
where we have used the fact that the noise in different baselines and frequency channels are uncorrelated. The resulting signal to noise ratio (SNR) will then be given by
\begin{equation}
{\rm SNR} = \frac{1}{\sigma_N} \frac{\int d^2U \int d \nu ~n_{\rm B}(U, \nu)~ S(\vec{U}, \nu)}{\int d^2 U \int d \nu ~n_{\rm B}(U, \nu)}.
\label{snr}
\end{equation}

The bandwidth $B_{\nu}$ of the observations is simply the frequency resolution $\Delta \nu_c$ times the number of frequency channels. Note that the above definition (\ref{snr}) of the SNR implies that we weight the visibility signal $S(\vec{U}, \nu)$ at individual baselines by the number $n_{\rm B}(U, \nu)$ of baselines. Since $n_{\rm B}(U, \nu) \propto \sigma^{-2}(U,\nu)$, we have simply weighed the visibilities according to inverse of the noise error.


\section{Results }
\label{res}

\subsection{21-cm signal pattern around the sources}
\label{tb_pat}

In this section, we describe the 21-cm brightness temperature distribution around different types of isolated source models, i.e.,   PopIII, Galaxy, Mini-QSO and HMXBs. We have already discussed the fiducial values of different parameters in these models in Section \ref{source_pro}.

We choose the fiducial redshift to be 15 for presenting our results. The favoured reionization models, after the recent release of constraints by the Planck team \citep{2015arXiv150201589P}, prefer a reionization history which begins at $z \sim 15$ \citep{Mitra15}. Hence one expects that the overlap of the brightness temperature patterns will not be significant at these redshifts. The corresponding frequency of observations will be $\sim 90$ MHz, well within the band of SKA1-low.

The upper panels of Figure \ref{visi_red} show the brightness temperature pattern as a function of radial distance from an isolated source for four different source models. We have shown results for three different coupling models $A$, $B$ and $C$ in each panel. For model $C$, the signal pattern in general can be divided into four prominent regions: (i) The signal is absent inside the central ionized \HII bubble of the source. (ii) The \HII region is followed by a region which is neutral and heated by X-rays. In this region, $\TK > \TCMB$ and thus the signal is seen in emission. (iii) The third is a strong absorption region which is colder than $\TCMB$ as the X-rays have not been able to penetrate into this region. (iv) Beyond the absorption region, the signal gradually approaches to zero as the $\lya$ coupling becomes less efficient far away from the source. In model $B$ the $\lya$ coupling is assumed to be efficient throughout and hence the fourth region is absent. In addition, we assume the heating to be efficient in model $A$, and hence both the third and fourth regions are absent in this case.

If we now look at the difference between different source models, we find that all the models produce a prominent \HII bubble around the source, irrespective of the coupling model. The size of the ionized bubble\footnote{We define ionized regions as those with ionization fraction larger than 0.5.} is significantly smaller for the PopIII model compared to the other three because the stellar mass in this model is considerably smaller than the others. In fact, if we compare these models for the coupling model $A$, we find that all the models are qualitatively similar showing a \HII region followed by a prominent emission region, the only difference being the size of the \HII region for the PopIII case.

If we now concentrate on the coupling models $B$ and $C$, we find that the Galaxy and HMXBs models do not show any substantial emission beyond the \HII region. The Galaxy model does not have any X-ray photons, hence the heating in the neutral / partially ionized IGM is negligible. The X-rays produced by the HMXBs, on the other hand, are mostly in the very high energy regime (hard X-rays) which have relatively smaller interaction cross-section. Hence these photons are not efficient in heating the medium. The presence of these hard X-rays, however, makes the transition from \HII region to the absorption region somewhat smoother for the HMXBs compared to the Galaxy model. The other two models, PopIII and Mini-QSO contain sufficient soft X-rays so as to produce a prominent emission region. The transition between the emission and the absorption regions is very smooth in these two models because of the same soft X-rays.


\subsection{Visibilities of the sources}
\label{visi_source}

If we ignore the contributions from the system noise and other astrophysical foregrounds, the strength of the 21-cm visibility signal around a single source depends on the source properties and the state of the IGM itself. The lower panels of Figure \ref{visi_red} show the expected visibilities for different source models and different coupling models for our fiducial redshift of 15. The properties of the sources are taken to be identical to that used in the upper panels. Note that we plot only the absolute value of the visibilities. We also show the expected system noise for SKA1-low assuming 1000 hours of observation time and a single frequency channel of width 50 kHz.

\begin{figure*} 
\begin{center}
\includegraphics[scale=0.7]{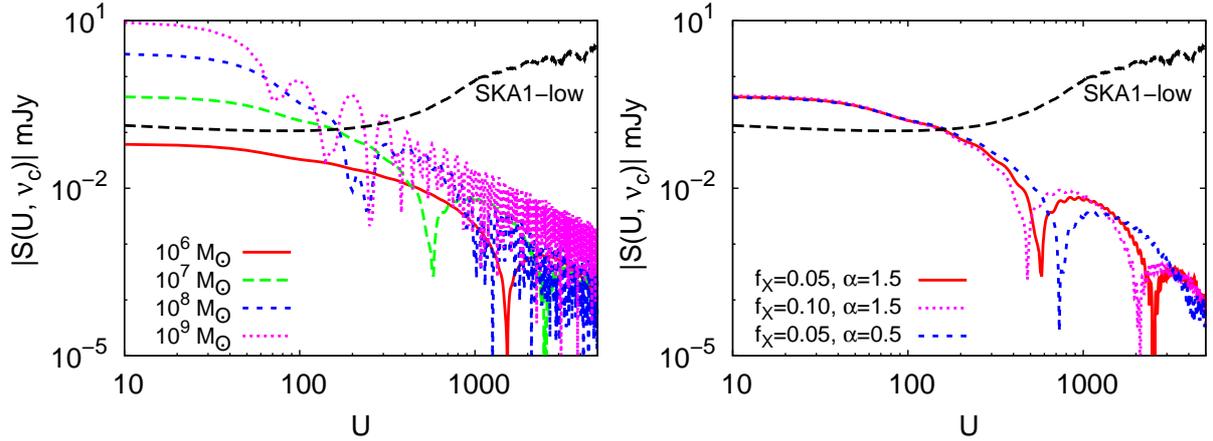}
    \caption{The dependence of the visibility on model parameters for the Mini-QSO source model. The coupling model is taken to be $C$. Left panel : The absolute visibility amplitudes for different values of stellar mass $M_{\star} = 10^{6}, 10^{7}, 10^{8}, 10^{9}\, \MSUN$ at redshift 15 as a function of baseline. All the other parameters are fixed to their fiducial values.  Right panel : The absolute visibility amplitudes as a function of baseline for different values of $f_X$ and $\alpha$.  In both the panels, the black dashed lines correspond to rms noise for 1000 h of observation with the SKA1-low with a frequency resolution of 50 kHz  and $\Delta U$=10 (see equation \ref{reduce_rms}).}
   \label{visifx}
\end{center}
\end{figure*}

The coupling model $A$ represents visibility of an ionized bubble around a source, similar to what was studied in \citet{kanan2007MNRAS.382..809D}. The signal is expected to show an oscillatory pattern arising from the spherical nature of the bubble (see appendix \ref{app_visi}). \citet{kanan2007MNRAS.382..809D} showed that the first  zero crossing of the visibility will occur at a baseline $U_0 = 0.61 r_\nu[R_b\sqrt{1-(\Delta \nu/\Delta \nu_b)^2}]^{-1}$, where $r_\nu$ is the comoving distance to the centre of the bubble (same as the $r(z)$ defined earlier), $R_b$ is the comoving radius of the \HII bubble, $\Delta \nu_b$ is the bubble size in frequency space and $\Delta \nu$ is the difference in frequency space between the centre of the bubble and the observed sky plane. For the source model PopIII, the size of the \HII region is $R_b \sim 0.23$ cMpc for which the first zero crossing occurs around $U_{0} \sim 28000$ , while the size is larger $R_b \sim 1.13$ cMpc for the other three source models which gives the first zero crossing at $U_0 \sim 5700$. These values are consistent with the prediction of \citet{kanan2007MNRAS.382..809D}. We also find that the visibility amplitude, which is proportional to $R^2_b$ at small $U$ (see Appendix \ref{app_visi} for details), is smaller for PopIII compared to the other three models. This too is because of the fact that the PopIII model produces a much smaller \HII region.

For models $B$ and $C$, the 21-cm signal visibilities are more complex and cannot be explained by a simple scenario as in model $A$. In general we find that the emission and absorption bubbles are larger than the \HII bubble for $B$ and $C$ respectively, and hence the first zero crossings occur at lower values of  $U$ compared to model $A$. For example, the first zero crossing in the source model Mini-QSO appears around $U_{0}=2390$ and 590 for models $B$ and $C$ respectively, while it appears around $U_0 = 5650$ for model $A$. In addition, we find that the amplitude of the visibility signal at small $U$ is the largest (smallest) for $C$ ($A$). The amplitude of the visibility at small baselines scales roughly as $\sim I_{\nu}~R_{b, \nu}^2$, where $I_{\nu}$ is the signal amplitude in the emission region for model $A$ and in the absorption region in models $B$ and $C$, and $R_{b, \nu}$ is the total radial extent (in the frequency channel $\nu$) of the ionized region for model A, ionized and emission regions for model B and  ionized, emission and absorption regions for models $C$. We have explained this aspect of the signal using simple models in Appendix \ref{app_visi}. Given this scaling, it is easy to see that since the size of the absorption region is much larger than the ionized and emission regions, the visibility amplitude in model $C$ would be the largest. This is further assisted by the fact that the $I_{\nu}$ itself is very high in the absorption region.

At this point, we can make some preliminary comment on the detectability of the signal with SKA1-low. In the bottom panels of Figure \ref{visi_red}, we have plotted the expected noise for a observing time of 1000 hours and for a frequency channel of width 50 kHz and $\Delta U = 10$. One can clearly see that the signal is detectable for smaller baselines $U \lesssim 100$ for model $C$ and for the source models Galaxy, Mini-QSO and HMXBs. The signal for the PopIII model, on the other hand, is substantially below the noise, and hence it is almost impossible to detect 21-cm signatures around a PopIII star. This result is consistent with other previous works, e.g., \citet{2014MNRAS.445.3674Y}. The signal for the coupling models $A$ and $B$ too are well below the noise leading to SNR $ < 1$, and hence detecting the first sources would be quite difficult once the $\lya$ coupling becomes efficient all throughout the IGM and/or the IGM is heated. We can also infer that detecting the first sources using the present telescopes (GMRT, LOFAR, MWA) will require unrealistically large amount of observing time as the noise amplitudes are at least an order of magnitude larger than the SKA1-low (see the right panel of Figure \ref{ska_base}). Keeping these in mind for the rest of the paper, we will consider the signal only from model $C$ and compare with SKA1-low noise levels. Also, for definiteness, we concentrate on the source model Mini-QSO, though many of the conclusions would hold for the models Galaxy and HMXBs.

Until now, we have presented our results for the fiducial values of the source and IGM parameters. To understand the dependence of the visibility signal on different parameters, we show in the left panel of Figure \ref{visifx} the absolute visibility for the source model Mini-QSO having different values of stellar mass. The coupling model used is $C$ and all other parameters have been set to their fiducial values. As expected, the absolute visibility increases with mass at small baselines.  In addition, the position of the first zero crossing shifts towards lower baselines as the mass increases which is because of the increase in the size of the 21-cm pattern around the source. We expect similar changes in the visibility signal when we increase the age of the source. The right panel of the figure shows the effect of different X-ray parameters on the signal visibility. We find that the visibility at small baselines is relatively insensitive to the parameters $f_X$ and $\alpha$. There is a slight change in the position of the first zero crossing for different $f_X$ and $\alpha$ because of the change in size of the heated regions. Also the signal at large baselines can be quite different when we change the parameters. However, the detectability of the signal, which is mostly driven by the smaller baselines, is expected to be relatively independent of the X-ray parameters.

\begin{figure*} 
\begin{center}
\includegraphics[scale=0.7]{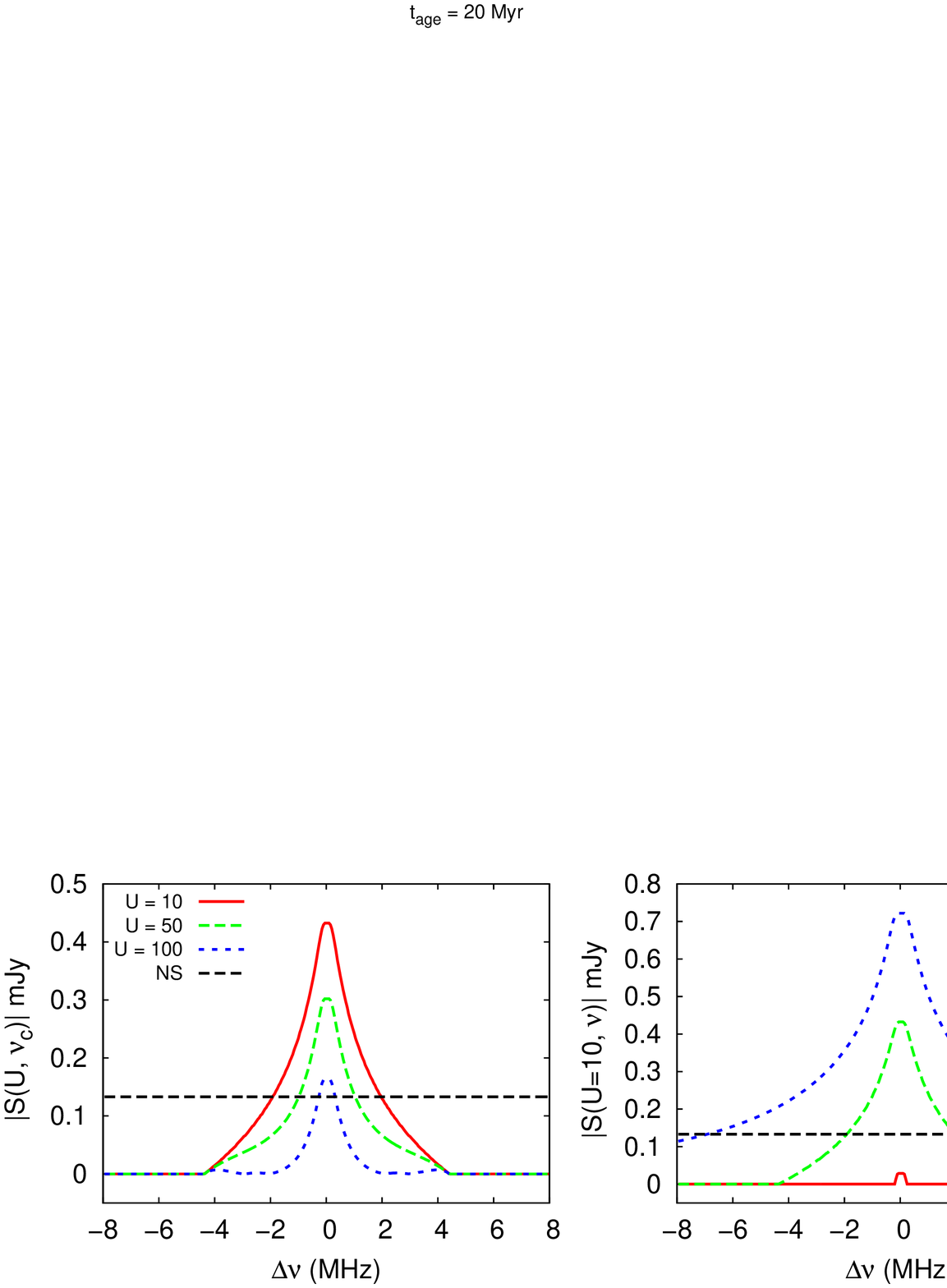}
    \caption{Left panel: The absolute visibility amplitude as a function of frequency channels at baselines $U=10, 50$ and 100 for the Mini-QSO source model and coupling model $C$. The parameters are fixed to their fiducial values. The quantity $\Delta \nu$ represents the difference between the frequency channel under consideration and the central frequency ($\nu_c$) of observation. Right panel : The absolute visibility amplitude as a function of frequency channels for different values of the source age $t_{\rm age} =  1, 20$ and 100 Myr at a baseline $U=10$.  The horizontal dashed lines in both panels represent the rms noise at baseline $U=10$  for 1000 h of observation time and 50 kHz frequency resolution with the SKA1-low.}
   \label{visinu1d}
\end{center}
\end{figure*}

Figure \ref{visinu1d} shows frequency dependence of the absolute visibility. The left panel shows the signal for different baselines when all the parameters are fixed to their fiducial values. The visibility peaks at the central channel at the position of the centre of the source and extends up to the size of the signal in the frequency space. As expected, the signal is larger for smaller baselines. The signal seems to increase more rapidly for smaller baselines than for larger ones. We found that the frequency dependence of the visibility for the Galaxy and HMXBs models is similar to that for the Mini-QSO model at small baselines. The amplitude of the absolute visibility increases with the age of the Mini-QSO, as shown in the right panel of Figure \ref{visinu1d}. We also see the extent of the signal in the frequency space increase with age as expected.

As we have seen, the detectability of the signal around a source is mostly determined by the smaller $U \lesssim 100$ baselines. At larger baselines, the signal amplitude decreases while the noise increases, thus making them unsuitable for detection. A possible way of characterizing the possibility of detection would be through integrating the signal over all baselines and frequency channels and define a SNR by using a inverse noise weighting as done in equation (\ref{snr}). The SNR we have defined contains contributions from all possible baselines and frequency channels. This may not be the optimum SNR possible for this scenario because, e.g., the SNR can be increased by neglecting large baselines and frequency channels away from the frequency channel that contains the source centre. However, constructing a optimum weighting scheme would require prior knowledge of the signal we are willing to detect. We rather concentrate on a robust estimate of the SNR which does not require any assumption about the nature of the signal around the source, which is more appropriate for the first detection of its kind.

\begin{figure*} 
\begin{center}
\includegraphics[scale=0.41]{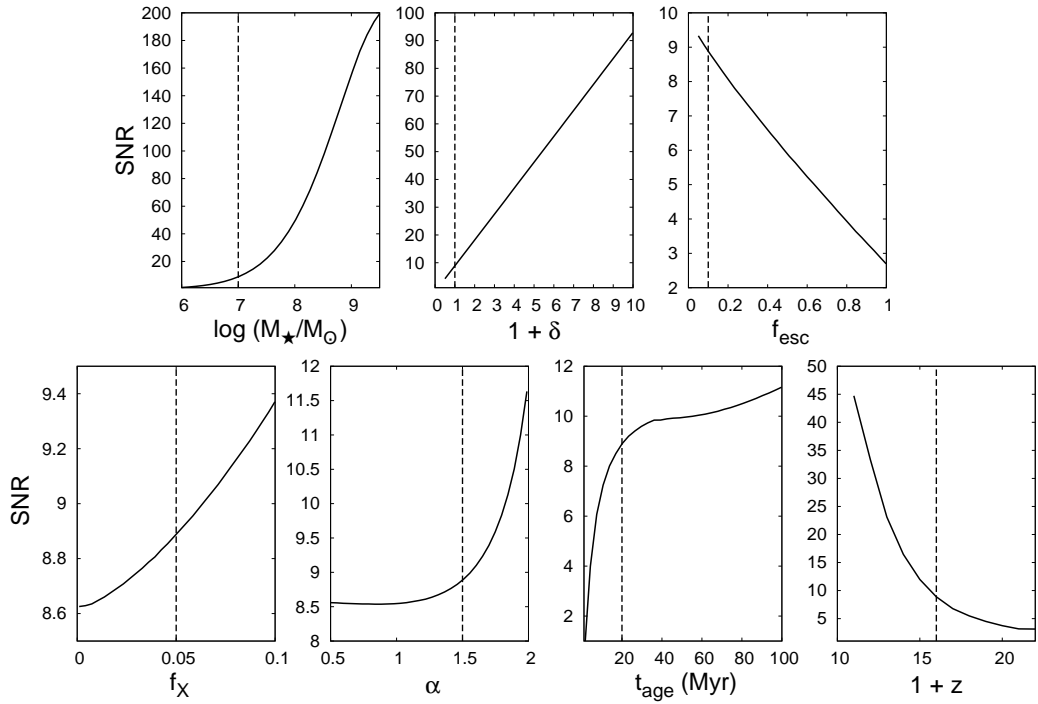}
    \caption{The SNR as a function of different model parameters for the source model Mini-QSO and coupling model $C$. The panels from top to bottom and left to right correspond to parameters stellar mass $M_{\star}$, overdensity $1+\delta$ of the medium around the source, escape fraction $f_{\rm esc}$ of ionizing photons, X-ray to UV luminosity ratio $f_X$, X-ray spectral index $\alpha$, age $t_{\rm age}$ of the source and redshift $z$. While calculating the dependence of the SNR on a particular parameter, we have fixed the other parameters to their fiducial values. The fiducial value for each parameter is denoted by the vertical dashed line in the corresponding panel. The SNR is calculated for 1000 hours of observation with SKA1-low with a bandwidth of 16 MHz. }
   \label{est1}
\end{center}
\end{figure*}

Different panels of Figure \ref{est1} show the SNR as a function of various parameters related to the source properties (e.g., stellar mass, escape fraction of ionizing photons, X-ray to UV luminosity ratio, spectral index and age of the source) and the overdensity of the surrounding IGM. Recall that the fiducial value of the parameters are $M_{\star} = 10^7\, \MSUN, \delta=0, f_{\rm esc}=0.1, f_X=0.05, \alpha=1.5, t_{\rm age }=20 $ Myr. While showing the dependence of the SNR on a particular parameter, we have fixed all the other parameters to their fiducial values. The characteristics of the observations are chosen such that the total observational time is 1000 hours, while the bandwidth is 16 MHz\footnote{We choose the bandwidth to be smaller than the total bandwidth of SKA1-low which is $\sim 300$ MHz. This is done to avoid contribution from other possible sources along the frequency direction.}.  As shown in the figure, the SNR is quite sensitive to the stellar mass $M_{\star}$, $f_{\rm esc}$, age $t_{\rm age}$ of the source and the IGM overdensity $1+\delta$. The SNR increases with the stellar mass (top left panel) as the strength of the 21-cm signal increases  and thus the visibility amplitude increases (see the left panel of Figure \ref{visifx}).  On the other hand, the radius of both the ionized and heated bubbles increase with the mass of the source, which result in faster decrease of the visibility with the baseline. Although the visibility signal is much stronger at smaller baselines for large stellar mass sources, the very small baselines $U_{\rm min} \lesssim 8$ are not available for SKA1-low baseline distribution at redshift 15.  This slows the growth of the SNR for the extremely high mass $M_{\star} \gtrsim 10^9 \MSUN$ sources. From the top right panel, we find that the SNR decreases with increase of escape fraction. This is because the amount of $\lya$ photons, produced due to the recombination of ionized hydrogen in the ISM, is proportional to $1 - f_{\rm esc}$ and thus the $\lya$ coupling becomes less effective as $f_{\rm esc}$ increases\footnote{A fraction $\sim 0.68$ of the ionizing photons absorbed in the ISM can be converted into $\lya$ photons because of the recombination of ionized hydrogen. Thus, in our calculation, we take the rate of $\lya$ photons, produced because of recombinations in the ISM, to be $0.68 (1-f_{\rm esc}) \dot{N}_{\rm ion}$ \citep{2014PASA...31...40D}, where $\dot{N}_{\rm ion}$ is the number of ionizing photons produced by the source per unit time.}. This leads to a smaller absorption region and thus a smaller amplitude of the 21-cm signal. The size of the 21-cm signal region increases with the age of the source as the photons are able to propagate longer distance and thus the visibility amplitude at lower baselines as well as the SNR increase with the age of the source, as shown in the bottom second right panel of Figure \ref{est1}. Since $\TB \propto 1 + \delta$ (see equation \ref{brightnessT}), the strength of the signal increases with the increase in the overdensity $1+\delta$. Hence the SNR increases almost linearly with $1+\delta$, as shown in the top middle panel of Figure \ref{est1}. The SNR in Figure \ref{est1} shows relatively weaker dependencies on the two X-ray parameters $f_X$ and $\alpha$.

\begin{table*}
\centering
\begin{tabular}{|l|c|c|c|c|c|c|c|}
\hline
Parameter $P$ & $M_{\star}$ & $1+\delta$ & $f_{\rm esc}$ & $f_X$ & $\alpha$ & $t_{\rm age}$ & $1+z$ \\
\hline
\hline
$P_{\rm fid}$ & $10^7 \MSUN$ & 1 & 0.1 & 0.05 & 1.5 & 20 Myr & 16 \\
\hline
$\gamma_P$ & 0.80 & 1.023 & $-0.42$ & 0.02 & 0.14 & 0.52 & $-3.69$ \\
\hline
applicable range & $1.3 \times 10^6$ -- $1.4 \times 10^8 \MSUN$ & 0.5 -- 10 & 0.05 -- 0.9 & 0.05 -- 0.1 & 0.5 -- 2 & 4 -- 100 Myr & 11 -- 24 \\
\hline
\end{tabular}
\caption[]{Dependence of the SNR on the parameters used in the study where the dependence is modelled as ${\rm SNR}(P) = 8.89 \left(P/P_{\rm fid}\right)^{\gamma_P}$. The final row of the table denotes the parameter range within which the fit produces the SNR to within $20\%$ of the actual value. The coupling model is taken to be $C$. The SNR is calculated for 1000 hours of observation with SKA1-low with a bandwidth of 16 MHz.}
\label{power_fit}
\end{table*}

Finally, we provide some scaling relations which would allow one to compute the SNR for a wide range of model parameter and redshift values. We model the relation as a simple power law
\begin{equation}
{\rm SNR}(P) = {\rm SNR}_{\rm fid} \left(\frac{P}{P_{\rm fid}}\right)^{\gamma_P},~~{\rm SNR}_{\rm fid} = 8.89
\label{snr_powerlaw}
\end{equation}
where $P$ is one of the seven parameters of interest, i.e., $M_{\star}$, $1+\delta$, $f_{\rm esc}$, $f_X$, $\alpha$, $t_{\rm age}$ and $1+z$. The fiducial value of the parameter $P$ is denoted as $P_{\rm fid}$, and the corresponding SNR is given by SNR$_{\rm fid}$. The power law index $\gamma_P$ determines the scaling relation for the parameter $P$ and the values are given in Table \ref{power_fit}. We have also indicated the values of the parameters where the scaling law reproduces the SNR to within 20\% of the actual value. It is possible to provide more accurate fits, however, those fits are more complex. Given the assumptions made in the physical modelling of the heating and ionized regions, the fit provided here should be sufficient to estimate the parameters which are favourable for detection. Note that the SNR also has dependence on the observational time as $t_{\rm obs}^{1/2}$. This is because of the fact that the SNR $\propto \sigma_{N}^{-1}$ (see equation \ref{snr}), which in turn depends on the observation time as $\sigma_N \propto t_{\rm obs}^{-1/2}$ (see equation \ref{snr_n}).

We can see from Table \ref{power_fit} that the dependence of the SNR on parameters $M_{\star}, f_{\rm esc}, t_{\rm age}$,  $1+\delta$ and $z$ are the most significant. In particular, we find that the SNR decreases sharply with increasing redshift which follows from the fact that $T_{\rm sys}$ increases rapidly at lower frequencies. For fiducial values of other parameters, the SNR would reduce from $\sim 9$ at $z=15$ to $\sim 3$ at $z=20$. We should mention here that, throughout the paper, we have calculated the rms noise for single polarization. The rms noise will decrease by a factor $\sqrt{2}$ for when both the polarizations are taken into account,  and therefore the SNR will improve by a factor $\sim 1.4$.



\begin{figure*}
\begin{center}
\includegraphics[scale=0.7]{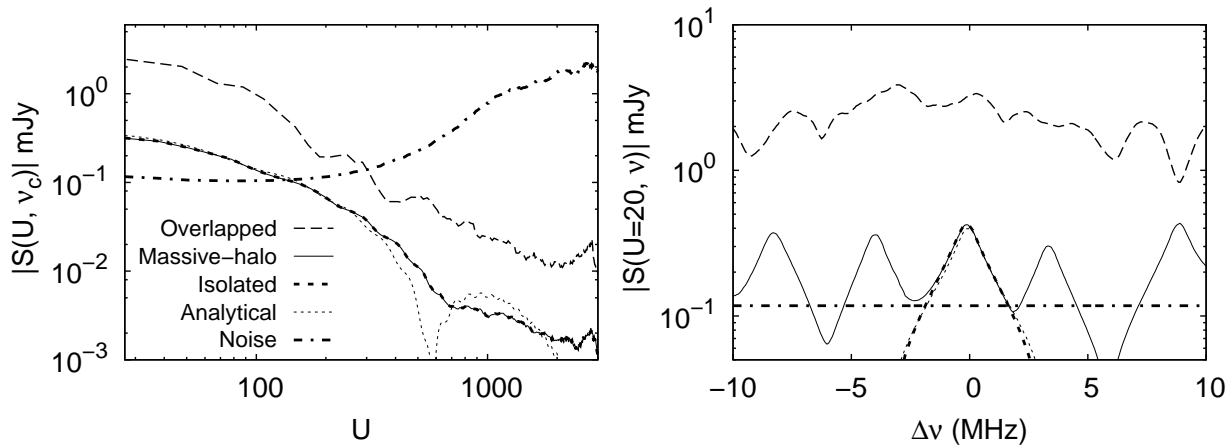}
    \caption{The absolute visibility amplitude calculated from the simulation box for different models. The model \emph{Isolated} (short-dashed) refers to the case where there is only one source in the box, while \emph{Massive-Halo} (solid) is the model where all haloes with masses $\gtrsim 2 \times 10^{10} \MSUN$ contain radiation sources. The model \emph{Overlapped} (long-dashed) corresponds to the case where all haloes in the box contain radiation sources and thus the individual 21-cm regions of individual sources overlap significantly. The model \emph{Analytical} (dotted) is simply the result from analytical calculation of the earlier sections. The coupling model is taken to be $C$. Left panel: The absolute visibility amplitude as a function of baseline. The dash-dotted curve represents the rms noise for SKA1-low for 1000 h of observation with a frequency resolution of 60 kHz. Right panel: The absolute visibility amplitude as a function of frequency channels at a baseline $U=20$.  The dash-dotted line represents the rms noise at a baseline $U=20$.}
   \label{mul_sour}
\end{center}
\end{figure*}

\subsection{Overlap between the sources}
\label{over_sou}

Till now we have been working under the assumption that there is only one source in the FOV for which the signal can be computed analytically.  However in reality there will be multiple sources in the FOV of interest and there could be overlap between the 21-cm regions of the individual sources. In addition, there could be effects arising from peculiar velocities along the line of sight. In such a scenario, it is almost impossible to find any analytical form  of the visibility and the only way we can compute the expected signal is by using a simulation box. To study such realistic scenarios, we have run a dark matter $N-$body simulation in a box of size 300 $h^{-1}$ cMpc with $2592^3$ particles. The mass resolution of the dark matter particles is $2 \times 10^{8}\, \MSUN$ and we are able to identify dark matter haloes down to $\sim 4\times 10^{9}\, \MSUN$ using the spherical overdensity method. We present our results for $z=15$ assuming coupling model $C$, the main reason being that the other models are not quite favourable for detection. The box length corresponds to a frequency bandwidth of 20 MHz at $z=15$ and the grid size gives a frequency resolution of $\sim 60$ kHz. Once we assign photon sources to the dark matter haloes, we use the method of \citet{ghara15a, ghara15b} to generate the brightness temperature maps. Note that the method naturally accounts for the effects of density fluctuations and the peculiar velocities of the gas in the IGM, as well as fluctuations in the spin temperature, while calculating the brightness temperatures maps. The method for generating the $\TB$ maps has been discussed in our earlier works \citep{ghara15a, ghara15b}. Here, we describe the method very briefly. 

\begin{itemize}
\item First we generate the $\XHII$ and $\TK$ profiles around sources for different values of the stellar mass, background gas density and redshift (as described in section \ref{1drt}), while the other parameters are kept fixed to their fiducial values.

\item After identifying the dark matter haloes using spherical over-density method in the simulation box, we assign the stellar mass to each halo. The relation between the stellar mass of a galaxy and the hosting dark-matter halo mass $M_{\rm halo}$ is assumed to be
\begin{equation}
M_{\star}=f_{\star} \left(\frac{{\Omega}_B}{{\Omega}_m} \right) M_{\rm halo},
\end{equation}
where $f_{\star}$ is the fraction of the baryon residing in stellar form in the galaxy.

\item We estimate the ionizing photons from each source and create ionized bubbles around them. In case there are overlaps between the bubbles, we calculate the number of unused ionizing photons and distribute them equally among the overlapping bubbles.

\item The hydrogen is partially ionized at the locations beyond the \HII regions of the sources. The ionized fraction $\XHII$ in these regions is estimated using the previously generated ionization profiles. We generate the  $\TK$ maps using a correlation of $\TK$ and $\XHII$ in the partially ionized regions \citep{ghara15a}.

\item We assume that the number density of the $\lya$ photons reduce as $1/R^2$, where $R$ is the distance from the center of the source. We simply add the $\lya$ flux from different sources to get the total $\lya$ flux at a region, which is later used to calculate the $\lya$ coupling coefficient. We can then generate the $\TB$ maps using the method as described in section \ref{1drt}.

\item We incorporate the peculiar velocity of the gas into the $\TB$ maps using the cell movement method of \citet{mao12}.

\end{itemize}

In order to test the results of the analytical results of the previous sections against the simulations, we first study a simple scenario where we assume that there is only one source in the box. We ensure that this source is at the centre of the FOV and calculate the visibility signal for the whole box. We fix the fraction of the baryon residing in the form of stars in the galaxy $f_{\star} = 0.005$. This is chosen such that the stellar mass of the source is $M_{\star} \sim  10^{7}\, \MSUN$.  We choose Mini-QSO source model and assume the age of the source to be 20 Myr. We choose the default values $f_X=0.05$ and $\alpha=1.5$ for the source. The comparison between this model, which we call \emph{Isolated}, and the analytical calculations is shown in Figure \ref{mul_sour}. In the left panel, we plot the absolute visibility for a single frequency channel of width 60 kHz as a function of baseline $U$, while in the right panel we show the same as a function of frequency channels for a baseline $U = 20$. One can see from the both the panels that the match between \emph{Isolated} and \emph{Analytical} is almost perfect. There is some deviation at larger baseline $U \gtrsim 300$ where the analytical results show sharp features while the simulation results are slightly smoothed out. This difference arises from the fluctuations in the density and velocity fields which are present in the simulations box (but do not exist in our analytical calculations). The resulting SNR, which is anyway dominated by the smaller $U \lesssim 100$ baselines, turns out to be very similar for the two cases.

We next study a more realistic model where the box contains more than one source. In order to ensure that there is no significant overlap between the 21-cm patterns around the sources, we populate only haloes which have masses larger than $\sim 4\times 10^{9}\, \MSUN$ with ionizing sources. As before, we choose $f_{\star} = 0.005$ so that the most massive source in the box has $M_{\star} \sim  10^{7}\, \MSUN$. The resulting heating fraction is $3 \times 10^{-5}$ and the ionization fraction is $1.4 \times 10^{-5}$.  The signal, labelled as \emph{Massive-halo}, is shown in Figure \ref{mul_sour}. One can see from the left panel that there is no difference between the \emph{Massive-halo} and \emph{Isolated} models as far as the $U$-dependence of the visibility in the central frequency channel is concerned. The main difference can be seen when we plot the signal as a function of the frequency channel as is done in the right panel. Clearly, the \emph{Massive-halo} shows multiple peaks arising from multiple sources in the FOV. However, the signal around each of the peak follows the prediction from the \emph{Analytical} model. In fact, the presence of such peak-like features in the visibility along the frequency axis could be a potential way of identifying the location of the sources in the frequency space\footnote{In principle, it should also be possible to locate the positions of the sources in the image plane using the observed visibilities. In general, if the source centre is offset by $\vec{\theta}_c$ from the centre of the field of view, there will be an additional phase shift of ${\rm e}^{2\pi \vec{U}.\vec{\theta}_c}$ in the observed visibilities. Thus the information on $\vec{\theta}_c$ can be obtained by measuring the phases of the visibility.}. The SNR obtained, by integrating over the full bandwidth of 20 MHz, turns out to be 34.7 for the \emph{Massive-halo} which is considerably higher than that in the \emph{Analytical} model (where the SNR is 7.5). This is simply because the signal is contributed by multiple sources. In case one is interested in focussing on signal from a single isolated source, one should integrate over a smaller bandwidth around the source. For example, if we reduce the bandwidth to 4 MHz, the SNRs for the \emph{Massive-halo} and \emph{Analytical} models come out to be very similar $\sim 16$.

For completeness, we study another model \emph{Overlapped} where all haloes in the box are assigned ionizing sources. In this case, the overlap between 21-cm pattern of individual sources is quite significant, and hence we do not expect any match with the analytical results. We can see from the left panel of Figure \ref{mul_sour} that the signal for the \emph{Overlapped} model is much higher than the other three models, even for a single frequency channel. This implies that even a channel of width 60 kHz contains multiple sources in this case unlike the other models. While studying the frequency dependence of the signal (right panel), we find that the visibility does not show any prominent peak-like structure as was seen in the other models. This clearly implies that the 21-cm pattern of the individual sources have overlapped significantly, and hence it may not be straightforward to extract the signature of a single source from the complex signal. In such a scenario, it is probably worth exploring lower frequencies (i.e., higher redshifts) where the overlap is expected to be less than what is found at $z=15$.

\begin{figure*}
\begin{center}
\includegraphics[scale=0.7]{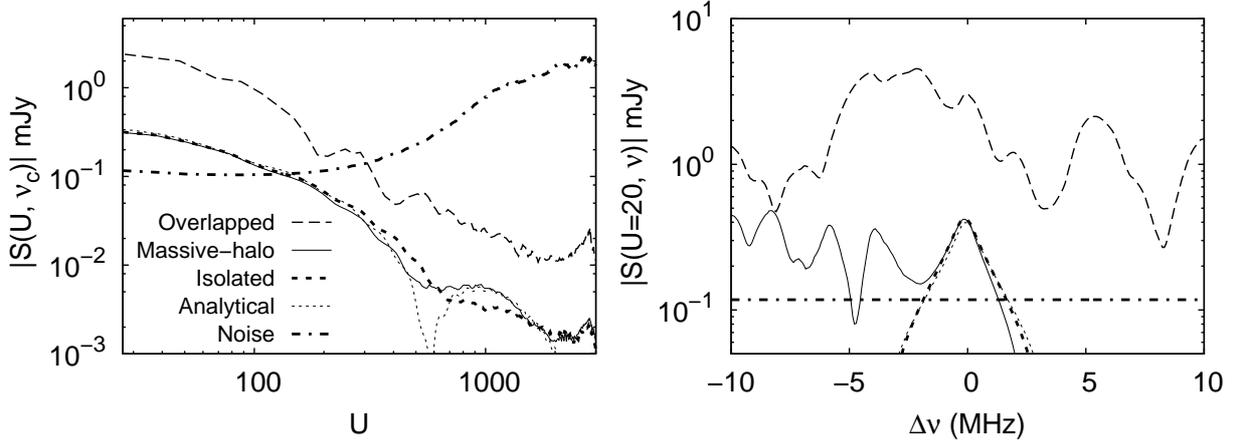}
    \caption{Same as Figure \ref{mul_sour}, but with the light-cone effect included.}
   \label{mul_sour_lc}
\end{center}
\end{figure*}

In addition, it is also important to account another line of sight effect while generating the redshifted 21-cm signal, namely the light-cone effect \citep{ghara15b}.  This effect is caused by the redshift evolution of the signal along the frequency direction. The detailed method of incorporating the light-cone effect in the 21-cm signal can be found in \citet{ghara15b}. While we expect the signal to be unaffected when there is only one source in the FOV, it can get modified significantly in more realistic scenarios where there are multiple sources in the FOV. On the average there will be less number of sources at the far end of the simulation box (i.e., at lower frequencies) along the line of sight direction. In order to study this effect we set the parameter $f_{\star}$ such that the stellar mass of the highest mass source at redshift 15 is $\sim 10^{7} ~\MSUN$.  The effect of the light cone is shown in Figure \ref{mul_sour_lc}, which should be compared with Figure \ref{mul_sour}. By construction, the signal at the central frequency channels look similar with and without the light-cone effect (as shown in the left hand side panels of the two figures) as both correspond to the same redshift 15. The main effect can be seen by comparing the right hand panels. For example, in the \emph{Massive-halo} scenario, there are less number of sources at channels with frequency smaller than the central frequency when the light-cone effect is included. On the other hand, there are more sources (and also the signal amplitude is higher) at channels with $\nu > {\nu}_c$ compared to the case where the light-cone effect was not considered. In the \emph{Overlapped} scenario, the amplitude of the signal is lower compared to the previous case (i.e., without light-cone effect) at smaller frequencies (far end of the simulation box) as the number density of sources is smaller. We see that the signal is also lower at the larger frequencies (near end of the box). This is because of the presence of larger number of sources in this region thus making the $\lya$ coupling relatively stronger. As a result the model $C$ approaches the model $B$ and gives rise to a signal visibility that is weaker compared to the case where light-cone effect was not considered. When the signal is integrated over the full bandwidth of 20 MHz, the SNR of the box including the light-cone effect is $\sim 48$. This is $\sim 40\%$ higher than the SNR without the light-cone effect. The main reason is that there are larger number of sources at the low redshift end of the simulation box. If we reduce the observational bandwidth to 4 MHz, the SNR is $\sim 15$, which is similar to the case without the light-cone effect.

We end this section by commenting upon identifying the possible location of the first sources. As long as the overlap between individual 21-cm patterns is not significant, it is possible to identify position of the source in the frequency space by plotting the visibility against $\nu$. The channels which show prominent peaks could be possible location of sources. A more direct method of identifying sources could be to use future NIR observations from an instrument like the JWST. It will probe the wavelength range of $0.6 - 29 \mu$m, which in principle can be used to detect the very first source in the universe. For example, the flux from a galaxy of stellar mass $M_{\star} = 10^7\, \MSUN$ at redshift 15, observed using the Near Infrared Camera (NIRCam),  is expected to be $\sim 5\times10^{-7} $ Jansky. This correspond to AB magnitude of 24.6, while the limiting magnitude of JWST is 29 \citep{2011ApJ...740...13Z}. These sources should be detected by the telescope with SNR larger than 10 for a observation time of $10^4$ second. 


\subsection{Effect of the astrophysical foregrounds}
\label{fore_deal}

We have assumed till now that the astrophysical foregrounds can be modelled perfectly and subtracted from the observed visibilities. In this section, we relax this assumption to some extent and investigate whether the signal can still be recovered. In general, an additional term $F(\vec{U},\nu)$ arising from the foregrounds needs to be added to the expression for the measured visibility, i.e., to the right hand side of equation \ref{visi_con}.

Astrophysical foregrounds, mostly contributed by the synchrotron radiation from our galaxy and extragalactic point sources, are generally several order stronger than the expected 21-cm signal. For the purpose of this analysis, the foreground contribution at each baseline and frequency channel is assumed to be a random variable with $\left\langle F(\vec{U}, \nu) \right\rangle=0$ (for all baselines except $\vec{U}=0$, which is anyway not considered in this study). The foreground contributions can be quantified in terms of the two-visibility correlation  $\left\langle F(\vec{U_1}, \nu{_1}) F(\vec{U_2}, \nu{_2}) \right\rangle$ given by \citep{kanan2007MNRAS.382..809D},

\begin{eqnarray}
\left< F(\vec{U_1}, \nu{_1}) F(\vec{U_2}, \nu{_2}) \right>  \!\!\!\! & = & \!\!\!\! \delta^{(2)}_{D}(\vec{U_1} + \vec{U_2}) \left( \frac{2 k_{ B}}{c^2} \right)^2 \left({\nu}_1 {\nu}_2 \right)^2 \nonumber\\
&\times& \!\!\!\! C_{2\pi U_1}({\nu}_1, {\nu}_2),
\label{equ_fore_ps}
\end{eqnarray}

where  $C_{l}({\nu}_1, {\nu}_2)$ is the multi-frequency angular power spectrum and the term $(2 k_B \nu^2/ c^2)$ is the conversion factor from brightness temperature to specific intensity. 

We consider three components of the foreground, namely, (i) the galactic synchrotron radiation, (ii) the Poisson noise arising from discrete point sources and (iii) the clustering contribution of the point sources. We assume that all these components have a power law dependence on the frequency \citep{2005ApJ...625..575S, kanan2007MNRAS.382..809D}. The angular power spectra of these components can be expressed as,

\begin{eqnarray}
C_{l}({\nu}_1, {\nu}_2) \!\!\!\! & = & A \left(\frac{{\nu}_f}{{\nu}_1} \right)^{\bar{\alpha}} \left(\frac{{\nu}_f}{{\nu}_2} \right)^{\bar{\alpha}} \left(\frac{1000}{l} \right)^{\beta} \nonumber\\
&\times& \!\!\!\! \exp{\left(-\log^{2}_{10} \frac{{\nu}_2}{{\nu}_1} \frac{1}{2{\xi}^2}\right)}
\label{equ_fore_anps}
\end{eqnarray}

We have taken the values of the parameters $A$, ${\nu}_f, \bar{\alpha}, \beta, \xi$ from \citet{kanan2007MNRAS.382..809D}, see their Table 1. While estimating the contribution from point sources we have assumed that all objects having flux larger than $5 \sigma$, where $\sigma$ is the rms noise in the high resolution continuum image, are identified and removed.

\begin{figure*}
\begin{center}
\includegraphics[scale=0.85]{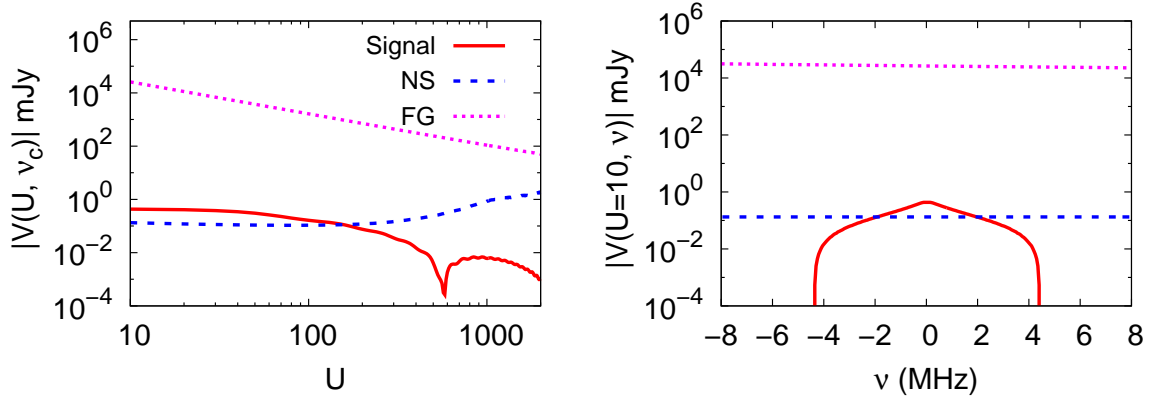}
    \caption{Left panel: The amplitude of the signal, the system noise and the foregrounds as a function of the baseline at the central  frequency channel. Right panel: The same quantities as a function of frequency difference from the central frequency of the observation at a baseline $U=10$.}
   \label{fore_1}
\end{center}
\end{figure*}

Figure \ref{fore_1} shows the contribution of the expected foreground for the SKA1-low for the instrument parameters listed in Table \ref{tab1}. Among the different components of the foreground considered in this study, the diffuse galactic synchrotron radiation is the most dominant one. The amplitude of the foreground is several orders larger than the expected signal and system noise for all relevant values of the baselines (left panel) and frequencies (right panel). Thus it is not possible to recover the signal by simply integrating over all baselines and frequency channels. One is hence compelled to use a different estimator, e.g., using a suitable filter \citep{kanan2007MNRAS.382..809D}, to recover the signal in presence of such a huge foreground contamination.

Unlike the redshifted 21-cm signal, the foregrounds have smooth frequency dependencies which is shown in the right panel of Figure \ref{fore_1}. This feature can be used to construct suitable filters to substantially reduce the foreground contributions. We essentially follow the method of \citet{kanan2007MNRAS.382..809D} for constructing such a filter $S_f(\vec{U}, \nu)$. The estimator $\hat{E}$ defined in equation (\ref{estimator-1}) can be modified to

\begin{equation}
\hat{E}= A_{\rm NS} \int {\rm d}^2 U \int {\rm d}\nu ~V(\vec{U}, \nu) ~S^{\star}_f(\vec{U}, \nu) ~{n}_{\rm B}(\vec{U}, \nu).
\label{equ_est}
\end{equation} 

Since the noise and the foregrounds are random numbers with zero mean, the expectation value of the estimator can be written as,
\begin{equation}
\left< \hat{E}\right>=A_{\rm NS}\int {\rm d}^2 U \int {\rm d}\nu ~S(\vec{U}, \nu) ~S^{\star}_f(\vec{U}, \nu) ~{n}_{\rm B}(\vec{U}, \nu)
\label{equ_exp_est}
\end{equation}
It can be shown that the noise  contribution is given by

\begin{eqnarray}
\left< (\Delta\hat{E})^2\right>_{\rm NS} \!\!\!\! & = & \!\!\!\! {{\sigma}^2_N} A_{\rm NS}  \nonumber\\  
&\times& \!\!\!\! \int {\rm d}^2 U \int {\rm d}\nu  ~|S_f(\vec{U}, \nu)|^2 ~{n}_{\rm B}(\vec{U}, \nu),
\label{equ_exp_est_NS}
\end{eqnarray}
while the contribution from the foregrounds are written as
\begin{eqnarray}
\left< (\Delta\hat{E})^2\right>_{\rm FG} \!\!\!\! & = & \!\!\!\! A^2_{\rm NS} \int {\rm d}^2 U \int {\rm d}{\nu}_1 \int {\rm d}{\nu}_2  \left( \frac{2 k_B}{c^2} \right)^2 \left({\nu}_1 {\nu}_2 \right)^2 \nonumber\\
&\times& \!\!\!\! {n}_{\rm B}(\vec{U}, {\nu}_1) ~{n}_{\rm B}(\vec{U}, {\nu}_2) ~C_{2\pi U}({\nu}_1, {\nu}_2) \nonumber\\
&\times& \!\!\!\! S^{\star}_f(\vec{U}, {\nu}_1)S_f(\vec{U}, {\nu}_2).
\label{equ_fore_NS}
\end{eqnarray}
The signal to noise ratio in presence of the foregrounds becomes
\begin{equation}
{\rm SNR} = \frac{\left< \hat{E}\right>}{\sqrt{\left< (\Delta\hat{E})^2\right>_{\rm NS} + \left< (\Delta\hat{E})^2\right>_{\rm FG}}}
\label{snr_est}
\end{equation}

\begin{figure*}
\begin{center}
\includegraphics[scale=0.7]{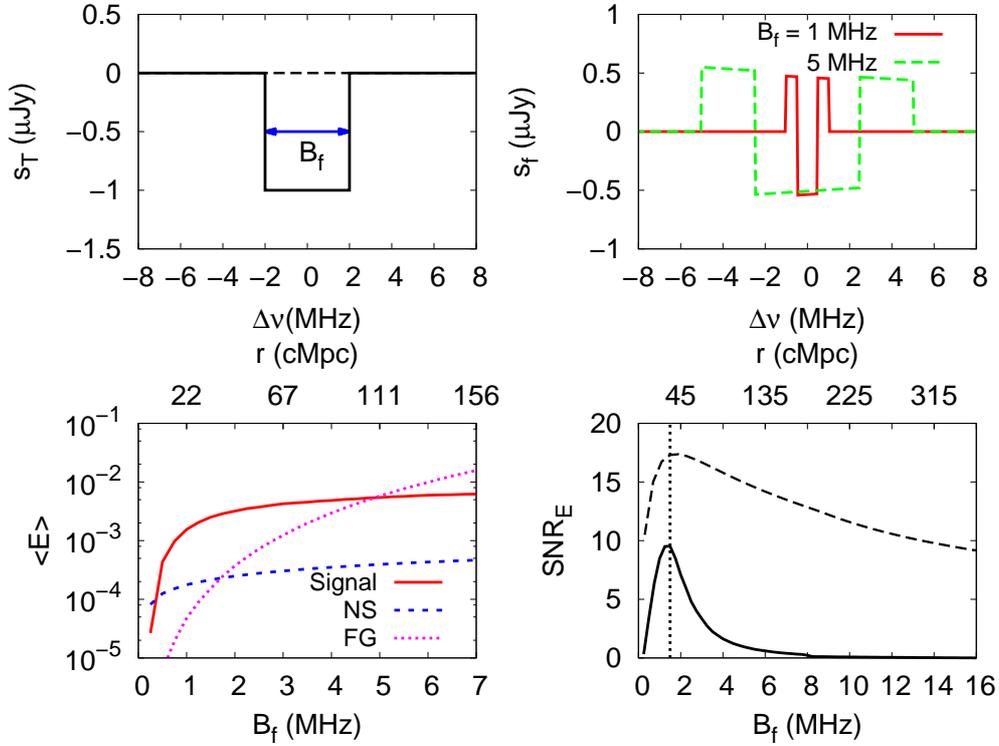}
    \caption{Top left: The bandpass filter function $S_T(\vec{U} , \nu, B_f)$ used for defining the filter used in this work (see equation \ref{equ_filter1}). Top right: The filter $S_f(\vec{U} , \nu, B_f)$ used in this work (see equation \ref{equ_filter}). Bottom left: The expectation value $\left<\hat{E}\right>$ of the estimator of the signal as a function of the parameter $B_f$. Also shown are the contributions of the system noise and the foregrounds quantified as the rms $\sqrt{\left<(\Delta \hat{E})^2\right>}$. The upper horizontal axis denotes the length scale corresponding to the band width $B_f$. Bottom right: Signal to noise ratio as a function of the parameter $B_f$ (solid line) in presence of the system noise and the foregrounds. The dashed line represents the foreground free case where the signal is integrated over a frequency bandwidth $B_f$. The vertical dotted line corresponds to the value of $B_f$ for which the SNR is maximum. The upper horizontal axis denotes the length scale corresponding to the band width $B_f$.}
   \label{fore_2}
\end{center}
\end{figure*}

The form of the filter $S_f$ is taken to be
\begin{eqnarray}
S_f(\vec{U}, \nu) \!\!\!\! & = & \!\!\!\! \left(\frac{\nu}{{\nu}_c} \right)^2 \left[  S_{T}(\vec{U}, \nu, B_{f}) - \frac{\Theta(1-\vert\nu - {\nu}_c\vert/B^{'})}{B^{'}}\right. \nonumber\\
&\times & \!\!\!\!\left. {\int}_{{\nu}_c-B^{'}/2}^{{\nu}_c+B^{'}/2}S_{T}(\vec{U}, {\nu}^{'}, B_{f}) ~{\rm d}{\nu}^{'}  \right], 
\label{equ_filter}
\end{eqnarray}
where $S_T(\vec{U}, \nu, B_f)$ is an ideal bandpass filter of width $B_f$ and is given by the rectangular function
\begin{eqnarray}
S_{T}(\vec{U}, \nu, B_f) \!\!\!\! & = & \!\!\!\! 0 \mbox{ if } |\nu - {\nu}_c|>\frac{B_f}{2} \nonumber\\
&=& \!\!\!\!  -1  \mbox{ if } |\nu - {\nu}_c|\leq \frac{B_f}{2}
\label{equ_filter1}
\end{eqnarray}
The $(\nu / \nu_c)^2$ term on the right hand side of equation (\ref{equ_filter}) accounts for the frequency dependence of the baseline distribution for a given array. The first term in the brackets on the right hand side ensures that the signal is integrated only over a frequency bandwidth of $B_f$ (rather than the full bandwidth $B_{\nu}$ of the observations), while the second term subtracts out the frequency-independent component in the range $\nu_c - B'/2$ to $\nu_c + B'/2$. The choice of $B'$ affects the results significantly. After some trial and error we choose $B^{'}=2 B_f$ if $B^{'}$ is less than $B_{\nu}$, else $B^{'}= B_{\nu}$. The advantage of the above filter is that it does not require any knowledge about the signal and the foregrounds, except for the fact that the foregrounds are smoother in the frequency. One can, in principle, improve the analysis by incorporating some knowledge of the signal into the filter (using, e.g., the so-called match filter technique), however, the purpose of this analysis is to show that the foregrounds can be dealt with even with simple filtering techniques.

The top left panel of the Figure \ref{fore_2} shows the frequency dependence of the bandpass function $S_{T}(\vec{U}, \nu, B_f)$. The top right panel shows the form of the filter $S_f(\vec{U}, \nu, B_f)$ for two different values of $B_f$. The bottom left panel shows the expectation value of the estimator $\left<\hat{E}\right>$ as a function of $B_f$. We also plot the contributions from the noise $\sqrt{\left<(\Delta \hat{E})^2\right>_{\rm NS}}$ and the foregrounds $\sqrt{\left<(\Delta \hat{E})^2\right>_{\rm FG}}$. The signal is calculated using the Mini-QSO model with fiducial parameters as in Figure \ref{visi_red}. We can see that the filter works quite efficiently in suppressing the foreground contributions for values of $B_f$ which roughly correspond to the size of the 21-cm pattern around the source. The contribution from the 21-cm signal saturates for larger values of $B_f$ while that for the foregrounds keeps on increasing. This suggests that the SNR should have a maximum for a $B_f$ corresponding to the size of the 21-cm pattern. The bottom right panel shows the dependence of SNR on $B_f$. The maximum SNR is found to be $\sim 9$, which corresponds to $B_f=1.5$ MHz and an equivalent length scale of $\sim 33$ cMpc. This length scale corresponds to the radius of the strongest absorption region as shown in Figure \ref{visi_red}. For comparison, we also show the SNR calculated when the foregrounds are not included in the analysis, and the signal is integrated over a bandwidth equal to $B_f$.  Clearly, the SNR is larger than that in presence of foregrounds, and approaches a value $\sim 9$ when $B_f \to 16$ MHz, consistent with results we obtained earlier.

We thus find that it should be possible to use the smooth frequency dependence of the foregrounds to suppress their contribution below the noise using appropriate filters. We obtain a SNR $\sim 9$ for our fiducial set of parameters using a simple filter that subtracts out the frequency-independent component. The analysis can be significantly improved by using detailed simulations of the foregrounds \citep{2006ApJ...650..529W, 2008MNRAS.391..383G, 2010MNRAS.409.1647J, Choudhuri2014MNRAS.445.4351C} and using more sophisticated subtraction techniques, e.g., \citep{2009ApJ...695..183B, 2010MNRAS.405.2492H, 2013ApJ...773...38G, 2015MNRAS.447..400A}. In the near future we plan to look into the effect of such advanced algorithms on the detectability of the 21-cm signal from the first stars.

\section{Summary and discussion}
\label{conc}

The main aim of this work is to predict the 21-cm visibility signal for different types of isolated sources at very high redshift and investigate the detectability of these sources using the SKA1-low. The source models we have considered are the PopIII stars, galaxies consisting of PopII stars, mini-quasars and HMXBs. We have used a one-dimensional radiative transfer code to track the time evolution of the ionization fractions of different ionization stage of hydrogen and helium as well as the kinetic temperature of the gas in the IGM. 

In general, the 21-cm signal around a source can be divided into four zones: (i) a central \HII region where the signal vanishes, (ii) followed by region where the signal is in emission, (iii) a strong absorption region just beyond the emission region, and finally (iv) the signal vanishes at very large distances from the source where the $\lya$ coupling is not efficient. Depending on the source model under consideration, one or more of these four regions may be non-existent or very small in size.

Our main findings can be summarized as follows:

\begin{itemize}

\item The signal from a PopIII star of mass $M_{\star} \sim 10^3 \MSUN$ will \emph{not} be detectable within any reasonable observing time with the SKA1-low. This is because the size of the region where the signal exists is too small, and so is the amplitude of the signal.

\item For other source models, i.e., normal galaxies, mini-quasars and HMXBs, the visibility signal is appreciable for small baselines $U \lesssim 100$. This indicates that the detection possibility can be increased by building large number of antenna elements in the core of the baseline design.

\item The detectability of the visibility signal is better for a model where the ionization, heating and $\lya$ coupling is determined by the radiation from the source itself (called model $C$ in this paper), i.e., there is no significant overlap between the signal patterns of individual sources. The overlap washes out the characteristics of the signal around a source, and in that case standard methods like estimating the power spectrum would be a better way to detect the signal. 

\item In order to maximize the detection prospects, we integrate the signal around the source over all baselines and a frequency bandwidth of 16 MHz. We find that it is possible to detect the signal around a source at $z=15$ with a SNR  $\sim 9$ in 1000 hours using the SKA1-low. This number is calculated for a source model where the galaxy contains standard stellar sources with stellar mass $M_{\star} = 10^7 \MSUN$ and age $t_{\rm age} = 20$ Myr. The escape fraction of the ionizing photons is $f_{\rm esc} = 0.1$. In addition to stellar sources, the galaxy is assumed to harbour a mini-quasar which produces X-rays with a power-law spectral index $\alpha = 1.5$ and a X-ray fraction $f_X = 0.05$. The surrounding IGM is assumed to have the mean cosmic density $\delta = 0$. The numbers are similar even when the mini-quasar is absent (i.e., no X-ray heating), or the X-ray is contributed mainly by HMXBs (i.e., almost no soft X-rays).

\item We also provide scaling relations which can be used for estimating the SNR for any other value of the parameters. We find that the SNR is quite sensitive to the stellar mass and age of the galaxy, as well as the escape fraction of ionizing photons. It is relatively less sensitive to the X-ray properties of the source. There is also a strong redshift dependence arising mainly from the fact that the system noise increases at lower frequencies.

\item We have verified our analytical calculations by comparing with the signal calculated using a more realistic simulation box which includes many sources. The visibility calculated from the simulation box is similar to the analytical calculations as long as the overlap between individual patterns is not significant. It is interesting to note that the visibility signal for smaller baselines as a function of frequency shows prominent peaks at channels where the source is present. This can potentially indicate channels around which the signal should be integrated.

\item While the galactic and the extragalactic foregrounds are several order higher than the signal we are trying to detect (and the system noise), it is possible to reduce the contribution of these by using suitable filters. We have shown that it is possible to achieve a signal to noise ratio $\sim 9$ even in the presence of the foregrounds by using filters which subtract out the frequency-independent component of the signal.

\end{itemize}

It is worth discussing some of the implications and caveats of our findings. For example, we have shown that the detection of the signal requires that the signal patterns of the individual sources do not overlap significantly, in particular, we require the $\lya$ coupling to become less efficient as we move away from the source. Now, this assumption can be violated in models where the $\lya$ background is set by a population of small mass $M_{\rm DM} \lesssim 10^8 \MSUN$ haloes. In that case one needs to look for signals either in objects with rather small stellar mass or in a scenario where model $B$ is the more appropriate one, both of which lead to smaller SNR. If we assume $f_{\star}$ to be $10 \%$, then a halo of $M_{\rm DM} \approx 3 \times 10^8 \MSUN$ produces a SNR of $\sim 5$ for model $C$, which probably is just above the detection threshold. For smaller mass haloes, one may need to implement more sophisticated detection techniques, e.g., stacking of the signal in source position, or a better weighing of the baselines while calculating the estimator. 

The analysis in this paper ignores various challenges which would arise while extracting the cosmological information. The presence of galactic and extragalactic foregrounds would contaminate the weak 21-cm signal from redshifted \HI, thus making the extraction a very difficult task. We have assumed in our work that the foregrounds have been subtracted perfectly from the data. Our conclusions need to be verified for the case where there is some residual foreground remains in the data which can give rise to spurious signals. The same holds true for other systematics, e.g., the calibration of the data against ionospheric instabilities. Given that we now have a good idea about the parameter space which is favourable for a detection with a very high SNR, it will worth exploring in the future these systematic effects on our analysis.

Another possible avenue to extend this work would be the understand how to determine the properties of the source once the signal is detected. This is important as the nature of the first stars still remains one of the most fundamental questions in contemporary cosmology. In this sense, our present work should be taken as a first step outlining the possibility of detecting the signal, which would eventually be followed up by understanding the source properties in detail.

\section*{Acknowledgement}

The authors would like to thank Prof. Mario Santos for providing us the proposed SKA baselines used in this study. We thank Rohit Sharma for providing us MWA baseline distribution. KKD would like to thank DST for support through the project SR/FTP/PS-119/2012 and the University Grant Commission (UGC), India for support through UGC-faculty recharge scheme (UGC-FRP) vide ref. no. F.4-5(137-FRP)/2014(BSR). We would like to thank the anonymous referee for providing constructive comments which helped us to improve the paper.

\bibliography{detectability_v4}

\appendix
\section{A simple model for calculating the visibility signal around isolated sources}
\label{app_visi}


\begin{figure*}
\begin{center}
\includegraphics[scale=0.7]{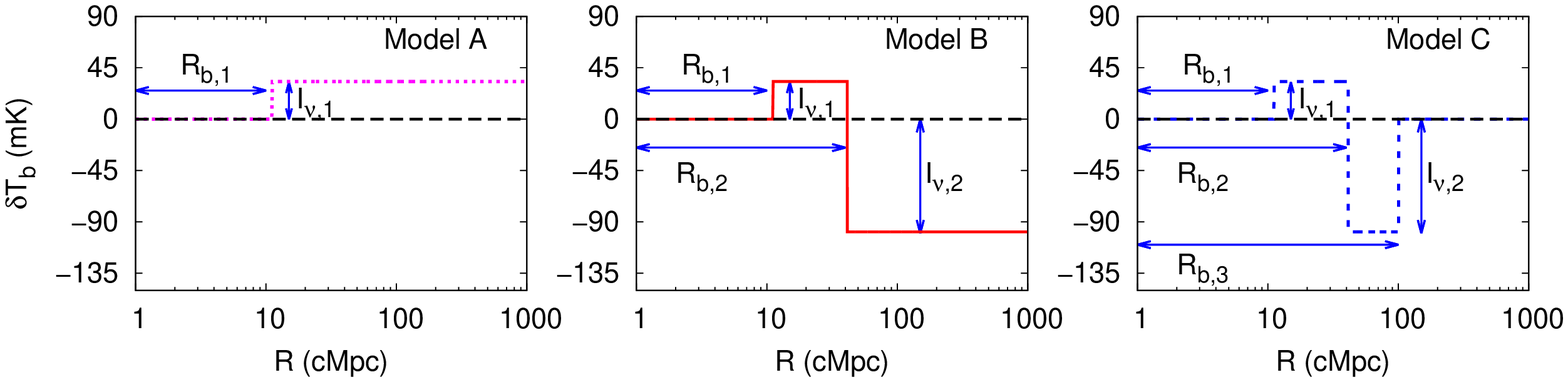}
    \caption{Differential brightness temperature as a function of the distance from the source for three simple scenarios. }
   \label{app_testtb}
\end{center}
\end{figure*}

\begin{figure*}
\begin{center}
\includegraphics[scale=0.9]{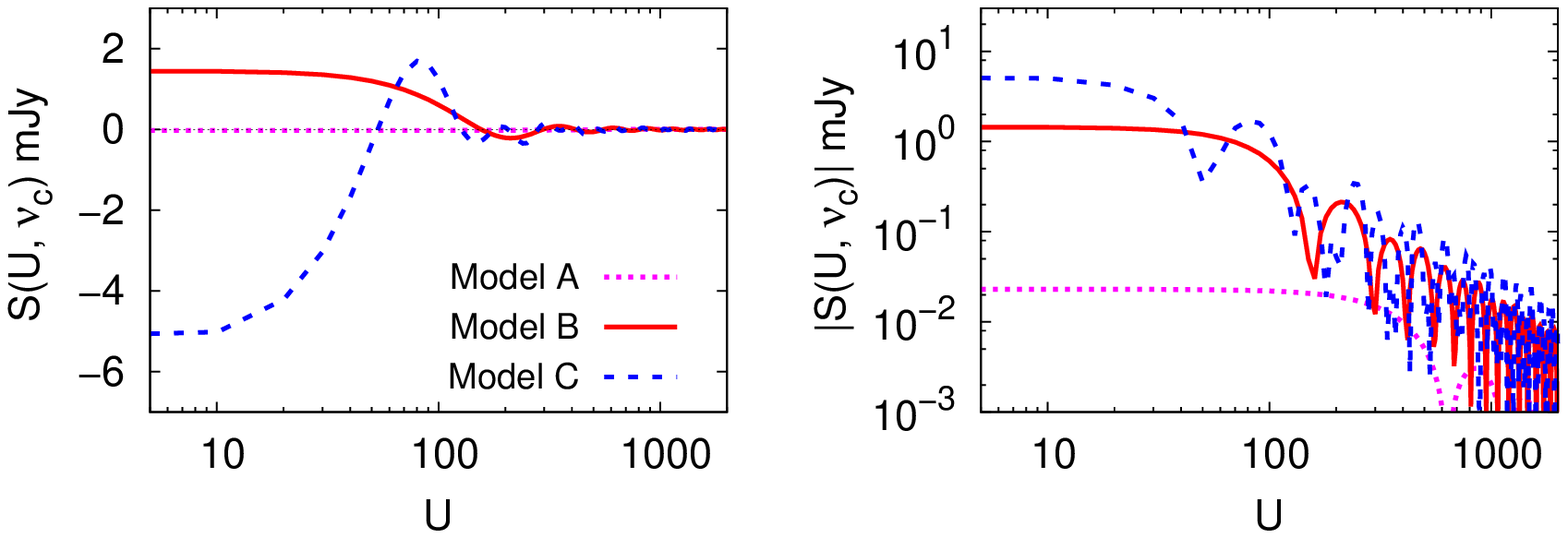}
    \caption{The visibility amplitude (left panel) and the absolute visibility (right panel) as a function of the baseline $U$ for three models $A$, $B$ and $C$ as shown in Figure \ref{app_testtb}. }
   \label{app_testvisi}
\end{center}
\end{figure*}

We present a simple model for calculating the 21-cm signal around a source which makes it easier to understand some of the properties. The main simplification is that we will approximate the signal in each zone, i.e., emission and absorption, by simple step-like functions of various sizes. For model $A$, where the $\lya$ coupling is efficient all over the IGM and the gas temperature is much larger than the CMBR temperature, the signal is zero within the \HII bubble and in emission uniform outside (assuming a uniform IGM). The signal can be represented by the shape shown in the top panel of Figure \ref{app_testtb}. \citet{kanan2007MNRAS.382..809D} showed that the visibility for $U > 0$ of a spherical bubble in the uniform IGM can be written as
\begin{equation}
S^{(A)}_{\rm c}(U,\nu)=-2\pi I_{\nu,1}\theta^2_{\nu, b,1} \left[\frac{J_1(2\pi U \theta_{\nu, b,1})}{2\pi U \theta_{\nu, b,1}}\right]\Theta\left(1-\frac{|\nu-\nu_c|}{\Delta\nu_{b,1}}\right),
\label{vis_bubble}
\end{equation} 
where $I_{\nu,1}$ is the 21-cm intensity in the neutral region, $\theta_{\nu,b,1}$ is the angular size of the bubble in the frequency channel $\nu$ and $\Delta\nu_{b,1}$ is the frequency extent of the bubble. The first zero crossing occurs at the baseline  $U_0^{(A)} = 0.61 / \theta_{\nu,b,1}$. Note that the angular extent in any frequency channel $\nu$ is related to that in the central frequency $\nu_c$ by $\theta_{\nu, b ,1} =  \theta_{b ,1} \sqrt{1-[(\nu - \nu_c)/\Delta \nu_{b,1}]^2}$.

We can similarly approximate the signal for model $B$ as that shown in the middle panel of Figure \ref{app_testtb}. We have assumed that both the emission and absorption regions have uniform intensity $I_{\nu,1}$ and $I_{\nu,2}$, with $I_{\nu, 2} < 0$. The positions of the transition between the regions are denoted as $\theta_{\nu, b, 1}$ and $\theta_{\nu, b, 2}$ respectively. In this case, the visibility for $U > 0$ is given by
\begin{eqnarray}
&&\!\!\!\!\!\!\!\!\!\!\!\!\! S^{(B)}_{\rm c}(U,\nu)=-2\pi I_{\nu,1}\theta^2_{\nu, b,1} \left[\frac{J_1(2\pi U \theta_{\nu, b,1})}{2\pi U \theta_{\nu, b,1}}\right]\Theta\left(1-\frac{|\nu-\nu_c|}{\Delta\nu_{b,1}}\right)
\nonumber \\
&&\!\!\!\!\!\!\!\!\!\!\!\!\! + 2\pi (I_{\nu,1} - I_{\nu,2})~\theta^2_{\nu, b,2} \left[\frac{J_1(2\pi U \theta_{\nu, b,2})}{2\pi U \theta_{\nu, b,2}}\right]\Theta\left(1-\frac{|\nu-\nu_c|}{\Delta\nu_{b,2}}\right).
\nonumber \\
\label{vis_bubble_2}
\end{eqnarray}
Note that since $I_{\nu,2} < 0$, the visibility amplitude for model $B$ is larger than that in model $A$ as can be seen in Figure \ref{app_testvisi}. In fact, for model $B$ we find $|I_{\nu, 2}| \gg I_{\nu, 1}$ for most source models, hence we can approximate the signal by
\begin{equation}
S^{(B)}_{\rm c}(U,\nu) \approx 2\pi |I_{\nu,2}|~\theta^2_{\nu, b,2} \left[\frac{J_1(2\pi U \theta_{\nu, b,2})}{2\pi U \theta_{\nu, b,2}}\right]\Theta\left(1-\frac{|\nu-\nu_c|}{\Delta\nu_{b,2}}\right).
\label{vis_B}
\end{equation}
Under this approximation, we find that the location of the first zero will be at $U_0^{(B)} \approx  0.61 / \theta_{\nu,b,2}$. Thus the extent of the signal in model $B$ is mostly determined by the position of the transition between emission and absorption regions. Also, we find that the position of the first zero is smaller for model $B$ than in model $A$, i.e., $U_0^{(B)} < U_0^{(A)}$. Since the amplitude $|I_{\nu,2}| \theta^2_{\nu, b,2} \gg I_{\nu, 1} \theta^2_{\nu, b,1}$, we see from the equations (\ref{vis_bubble}) and (\ref{vis_B}) that the amplitude of the signal at smaller baselines would be larger for model $B$ than for $A$.

Extending the above calculations to model $C$, as shown in the bottom panel of Figure \ref{app_testtb}, we find the visibility to be
\begin{eqnarray}
&&\!\!\!\!\!\!\!\!\!\!\!\!\! S_{\rm c}(U,\nu)=-2\pi I_{\nu,1}\theta^2_{\nu, b,1} \left[\frac{J_1(2\pi U \theta_{\nu, b,1})}{2\pi U \theta_{\nu, b,1}}\right]\Theta\left(1-\frac{|\nu-\nu_c|}{\Delta\nu_{b,1}}\right)
\nonumber \\
&&\!\!\!\!\!\!\!\!\!\!\!\!\! + 2\pi (I_{\nu,1} - I_{\nu,2})~\theta^2_{\nu, b,2} \left[\frac{J_1(2\pi U \theta_{\nu, b,2})}{2\pi U \theta_{\nu, b,2}}\right]\Theta\left(1-\frac{|\nu-\nu_c|}{\Delta\nu_{b,2}}\right)
\nonumber \\
&&\!\!\!\!\!\!\!\!\!\!\!\!\! + 2\pi I_{\nu,2}~\theta^2_{\nu, b,3} \left[\frac{J_1(2\pi U \theta_{\nu, b,3})}{2\pi U \theta_{\nu, b,3}}\right]\Theta\left(1-\frac{|\nu-\nu_c|}{\Delta\nu_{b,3}}\right).
\label{vis_bubble_3}
\end{eqnarray}
In this case too, let us make the assumption that $|I_{\nu, 2}| \gg I_{\nu, 1}$, which is reasonable for all the source models except PopIII. In addition, one can also make the assumption that $\theta_{\nu, b, 3} \gg \theta_{\nu, b, 2}$, i.e., the extent of the absorption zone is much larger than that of the emission. This too is a reasonable assumption (except for PopIII) as can be seen from the top panels of Figure \ref{visi_red}. In that case, the signal takes a simpler form
\begin{equation}
S^{(C)}_{\rm c}(U,\nu) \approx - 2\pi |I_{\nu,2}|~\theta^2_{\nu, b, 3} \left[\frac{J_1(2\pi U \theta_{\nu, b,3})}{2\pi U \theta_{\nu, b,3}}\right]\Theta\left(1-\frac{|\nu-\nu_c|}{\Delta\nu_{b,3}}\right).
\label{vis_C}
\end{equation}
In this case, the signal at small baselines would larger than those in model $B$ by a factor $\sim \theta^2_{\nu, b, 3} / \theta^2_{\nu, b, 2}$. Also, the location of the first zero would be $U_0^{(C)} \approx  0.61 / \theta_{\nu,b,3}$, which is smaller than that found in models $B$ and $A$.

We thus find, with the help of the simplified models and by making reasonable approximations, that the first zero of the visibility signal for models $A$, $B$ and $C$ are determined by the extent of the ionized bubble, emission region and absorption region respectively. The strength of the signal at small baselines (i.e., those smaller than the location of the first zero) would be determined by a combination of the intensities and region sizes.

\label{lastpage}
\end{document}